\documentclass[useAMS,usenatbib]{mn2e}
\usepackage{graphicx}
\usepackage{natbib}
\usepackage{amsmath}
\usepackage{float}
\usepackage{rotating}
\bibpunct{(}{)}{;}{a}{}{,}

\newcommand{\ie}{$i.e.,\;$}
\newcommand{\eg}{$e.g.,\;$}

\usepackage{txfonts}

\title[]{Kiloparsec-scale radio emission in Seyfert and LINER galaxies}
\author[Singh et al.]{Veeresh Singh$^{1,2,3}$\thanks{E-mail: Singhv4@ukzn.ac.za}, 
C. H. Ishwara-Chandra$^{1}$, Yogesh Wadadekar$^{1}$, Alexandre Beelen$^{2}$ \newauthor and Preeti Kharb$^{4}$ \\
$^{1}$National Centre for Radio Astrophysics, TIFR, Post Bag 3, Ganeshkhind, Pune 411007, India\\ 
$^{2}$Institut d'Astrophysique Spatiale, B$\hat{\rm a}$t. 121, Universit{\'e} Paris-Sud, 91405 Orsay Cedex, France \\
$^{3}$Astrophysics and Cosmology Research Unit, School of Chemistry and Physics, University of KwaZulu-Natal, Durban 4041, South Africa \\
$^{4}$Indian Institute of Astrophysics, Bangalore 560034, India \\}

\begin{document}

\date{Accepted ........; Received .........; in original form ........}

\pagerange{\pageref{firstpage}--\pageref{lastpage}} \pubyear{2014}

\maketitle

\label{firstpage}

\begin{abstract}

Seyfert and LINER galaxies are known to exhibit compact radio emission on $\sim$ 10 to 100 parsec scales, but larger Kiloparsec-Scale Radio 
structures (KSRs) often remain undetected in sub-arcsec high resolution observations.
We investigate the prevalence and nature of KSRs in Seyfert and LINER galaxies using 
the 1.4 GHz VLA FIRST and NVSS observations. 
Our sample consists of 2651 sources detected in FIRST and of these 1737 sources also have NVSS counterparts.
Considering the ratio of total to peak flux density ($\theta$ $=$ ${\rm (S_{\rm int}/S_{\rm peak})^{1/2}}$) as a parameter to infer the presence of 
extended radio emission we show that $\geq$ 30$\%$ of FIRST detected sources possess extended radio structures on scales larger 
than 1.0 kpc. 
The use of low-resolution NVSS observations help us to recover faint extended KSRs that are resolved out in FIRST observations and results in 
$\geq$ 42.5$\%$ KSR sources in FIRST$-$NVSS subsample. 
This fraction is only a lower limit owing to the combination of projection, resolution and sensitivity effects.
Our study demonstrates that KSRs may be more common than previously thought and are found across all redshifts, luminosities and radio-loudness. 
The extranuclear radio luminosity of KSR sources is found to be positively correlated with the core radio luminosity as well as the 
[O~III] $\lambda$5007{\AA} line luminosity and this can be interpreted as KSRs being powered by AGN rather than star$-$formation. 
The distributions of the FIR$-$to$-$radio ratios and mid$-$IR colors of KSR sources are also consistent with their AGN origin. However, 
contribution from star$-$formation cannot be ruled out particularly in sources with low radio luminosities.

\end{abstract}


\section{Introduction}
Active Galactic Nuclei (AGN) are classified into radio$-$loud and radio$-$quiet subclasses based on the ratio of 5 GHz radio flux density 
to 4400{\AA} optical continuum flux (R $=$ $\frac{{{\nu}_{\rm 5 GHz}}{\rm S_{\rm 5 GHz}}}{{\nu}_{\rm 4400{\AA}}{\rm S_{\rm 4400 {\AA}}}}$) 
with R $\geq$ 10 for radio-loud AGN and R $<$ 10 for radio-quiet AGN \citep{Kellermann89}. 
This dichotomy may also be manifested in radio morphologies as radio-loud AGN exhibit bipolar jets terminating into radio lobes 
with total end-to-end size spanning upto few hundred kiloparsec to even few Megaparsec. 
While, radio-quiet AGN often display compact nuclear radio emission with occasional presence of extended radio emission 
limited to $\leq$~10 kiloparsec scale. 
Seyfert and Low Ionization Nuclear Emission Region (LINER) galaxies are classified as radio$-$quiet AGN with 
low optical luminosity (M$_{\rm B-Band}$ $>$ -22.25 assuming H$_{0}$ = 71 km s$^{-1}$ Mpc$^{-1}$) \citep{Schmidt83}. 
High resolution radio observations show that most of the Seyfert galaxies possess compact sub-parsec scale nuclear 
emission with brightness temperature $\geq$ 10$^{7}$ K \citep{Thean2000,Middelberg04,Lal04}.
However, radio observations of arcsec resolution have revealed resolved structures, with hints of jets and extended emission in several 
Seyfert galaxies \citep{Baum93,Colbert96,Morganti99,Gallimore06}. 
This is further supported by the examples of few individual Seyfert galaxies that exhibit radio morphology consisting of a core, collimated jets 
and lobes similar to those found in radio-loud AGN ({\eg}NGC 1052 \citep{Wrobel84}, NGC 1068 \citep{Ulvestad87}, 
NGC 7674 \citep{Momjian03}, MRK 3 \citep{Kukula99} and MRK 6 \citep{Kharb06}), although, unlike radio-loud AGN the linear radio structures 
in Seyfert galaxies span only upto few kiloparsec. 
Also, Kpc-Scale Radio structures (KSRs) in Seyfert galaxies are not always found to be aligned with the pc-scale jet-like structures 
\citep{Colbert96,Kharb06,Kharb10}. 
It is believed that the low-power radio jets seen in Seyfert galaxies are analogous to the larger jets seen in radio galaxies but 
perhaps distorted or stunted by the interaction with the surrounding inter-stellar medium (ISM) of the host galaxy 
\citep{Whittle04,Gallimore06}. 
Overall it appears that the radio structures in Seyfert galaxies are much more complex than in powerful radio-loud AGN. 
\par
It is important to note that most of the radio studies of Seyfert/LINER samples have been carried out mainly at higher frequencies 
with long-baseline interferometers ({\eg}VLA A-array, MERLIN, VLBI) that filter out low-surface-brightness emission distributed over larger angular 
scales \citep[{e.g.,}][]{Ulvestad89,Kukula95,Thean2000,Ho01,Lal04,Nagar05,Panessa13}. 
Therefore, high resolution observations are likely to miss the detection of KSRs.
Indeed, in some Seyfert galaxies, the radio flux density arising from their parsec-scale structure is found to be much lower than 
that derived from observations with lower resolution, even if the nucleus appears unresolved. 
The missing flux has been interpreted as due to the presence of KSRs of low-surface-brightness \citep{Sadler95,Orienti10}. 
In fact, sensitive radio observations of arcsec resolution have suggested that there may be a large fraction of 
Seyferts with KSRs \citep[{\eg}][]{Baum93,Kukula96,Gallimore06}.  
Using deep VLA `D' array observations ($\sim$ 15 - 20 arcsec resolution at 5 GHz with average rms $\sim$ 50~$\mu$Jy) \cite{Gallimore06} 
showed that $\geq$ 44$\%$ sources in their sample of 43 Seyfert and LINER galaxies exhibit extended radio structures at 
least 1 kpc in total extent that do not match the morphology of the disk or its associated star-forming regions.
This favors a scenario in which KSRs are likely to be an extension of the nuclear radio jet and therefore related 
to AGN activity. However, starburst galaxies ({\eg}M82 \citep{Seaquist91} and NGC 253 \citep{Carilli92}) are also known to 
display KSRs probably originating from a superwind generated by the cumulative effects of stellar winds. 
These results raise questions whether KSRs are truly so common in Seyfert and LINER galaxies and 
whether they are related to the AGN or starburst.
We note that all the previous studies on KSRs are limited to targeted observations of fairly small samples despite the availability of 
large Seyfert/LINER samples.  
In order to detect and examine the prevalence and nature of KSRs in Seyfert and LINER galaxies 
we study radio emission properties in a large sample of $\sim$ 2651 sources using 1.4 GHz VLA FIRST and NVSS radio surveys. 
Our sample of Seyfert and LINER galaxies is derived from the 13th edition AGN catalog of \cite{Veron-Cetty10} and 
is the largest sample of Seyfert and LINER galaxies hitherto used for KSRs studies.
\par
This paper is organized as follows. In Section 2, we discuss our sample selection and its merits. 
The details on the radio data are given in Section 3 and the identification of FIRST and NVSS radio counterparts 
is discussed in Section 4. The methods of using flux densities to ascertain the presence of extended emission are discussed in Section 5. 
A brief discussion on the sources with multicomponent radio structures is presented in Section 6. 
Section 7 is devoted to the discussion on the origin of KSRs. 
The conclusions of our study are given in Section 8.    
\\
In this paper we have assumed H$_{0}$ = 71 km s$^{-1}$ Mpc$^{-1}$, ${\Omega}_{\rm M}$ = 0.27, ${\Omega}_{\Lambda}$ = 0.73.

\section{The sample}
Our sample of Seyfert and LINER galaxies is derived from the 13th edition AGN catalog of \cite{Veron-Cetty10} (hereafter VV10) 
which is the most updated complete survey of the literature and lists all the quasars, BL Lacs, active galaxies known to the catalog 
authors prior to July 1, 2009.
The VV10 catalog follows conventional definition of Seyfert and LINER galaxies that are defined as AGN of low optical luminosity 
with absolute B-band magnitude fainter than -22.25 ({\ie}M$_{\rm B}$ $>$ -22.25) \citep{Schmidt83}. 
The use of Hubble parameter H$_{0}$ = 71 km s$^{-1}$ Mpc$^{-1}$ instead of 50 km s$^{-1}$ Mpc$^{-1}$ has changed the 
optical luminosity limit to M$_{\rm B}$ $=$ -22.25 rather than M$_{\rm B}$ $=$ -23.0 considered in some early studies. 
The VV10 catalog contains total 34231 Low Luminosity AGN (LLAGN) with M$_{\rm B}$ $>$ -22.25. 
To select our sample of Seyfert and LINER galaxies we have taken only those sources that have confirmed Seyfert or LINER classification in 
the VV10 catalog and excluded sources without classification or with uncertain classification. 
We refer Seyfert and LINER galaxies collectively as `LLAGN'. 
We obtain a sample of 23448 LLAGN that contains 16517 Seyfert 1s (designated as `S1'), 6024 Seyfert 2s (designated as `S2'), 
and 907 LINERs (designated as `S3'). 
In our sample we have grouped intermediate Seyferts ({\ie}types 1.0, 1.2, 1.5, 1.8, and 1.9) into Seyfert type 1s 
noting the fact that intermediate Seyferts are classified based on the appearance of broad permitted Balmer lines which are characteristics of 
type 1s \citep{Osterbrock81}.  
Narrow line Seyfert galaxies (designated as `S1n') and Seyfert galaxies with broad permitted lines detected in infrared wavelengths 
(designated as `S1i') or in polarized light (designated as `S1h') are also included in Seyfert type 1s.
\par
\begin{figure}
\includegraphics[angle=0,width=9.2cm]{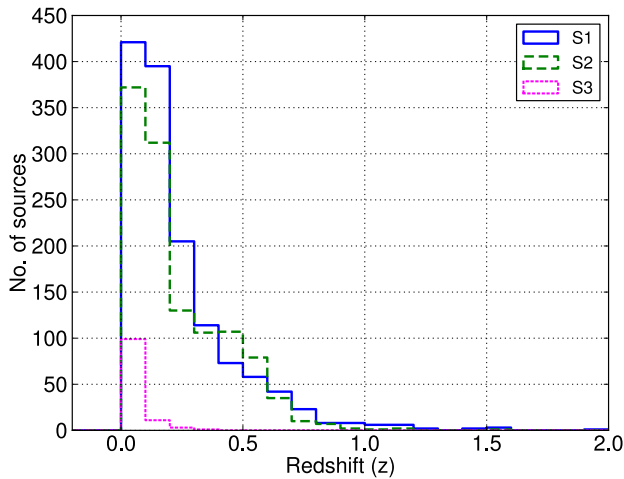}
\caption{Redshift distributions of Seyfert 1s, Seyfert 2s and LINERs (S3) detected in FIRST.}
\label{fig:RedshiftHist}
\end{figure}

Table~\ref{table:LLAGNNumbers} lists the number of Seyferts and LINERs extracted from the VV10 catalog and their detections in FIRST and NVSS radio surveys.  
Figure~\ref{fig:RedshiftHist} shows the redshift distributions of all Seyfert and LINER galaxies detected in FIRST survey.  
The FIRST detected LLAGN have median redshift ($z_{\rm median}$) $\sim$ 0.15, with 98.7$\%$ at $z~\leq~1.0$. 
In comparison to Seyfert galaxies, LINERs are seen at lower redshifts with median redshift ($z_{\rm median}$) $\sim$ 0.048. 
The FIRST detected Seyfert type 1s and type 2s have similar redshift distributions with median redshifts $\sim$ 0.158 and $\sim$ 0.162, respectively.
Two sample Kolmogorov-Smirnov (KS) test shows that the redshift distributions of Seyfert 1s and Seyfert 2s are similar (D $\sim$ 0.05) 
with probability of null hypothesis that the two samples are drawn from two different parent population is only 0.095.  
Since our final sample consists of FIRST detected sources ({\ie}radio selected) and radio emission is optically thin to dust obscuration, 
therefore, our sample is unaffected by the biases introduced by obscuration caused by dusty torus present around AGN. 
This is evident from the comparison of redshift distributions of Seyfert type 1s and type 2s. 
In general, optically selected Seyfert samples are known to be biased against Seyfert type 2s as AGN view is intercepted by the 
circumnuclear obscuring torus \citep{Ho01}. 

\begin{table}
\centering
\begin{minipage}{140mm}
\caption{Number of Seyferts and LINERs detected in FIRST and NVSS}
\begin{tabular}{@{}cccc@{}}
\hline
Class                    &  S1 & S2 & S3  \\ \hline
No. of sources in VV10 catalog   &    16517   & 6024  &  907    \\
No. of sources detected in FIRST &  1373  &  1164  &  114     \\
No. of sources detected in both FIRST and NVSS &  884   & 753  & 100   \\
\hline
\end{tabular}
\label{table:LLAGNNumbers} 
\end{minipage}
\end{table}

\section{Radio data}
We use the ``{\em Faint Images of the Radio Sky at Twenty-cm}'' (FIRST\footnote{http://sundog.stsci.edu/}) 
survey carried out at 1.4 GHz by Very Large Array (VLA) in its B-configuration and it 
covers sky region over 10,000 square degrees of the North and South Galactic Caps \citep{Becker95}.
The FIRST radio maps have resolution of $\sim$ 5$\arcsec$ and typical rms down to $\sim$ 0.15 mJy. 
We have used most recent version of FIRST source catalog (released on March 2014) that lists peak as well as integrated flux densities 
and radio sizes derived by fitting a two-dimensional 
Gaussian to a source detected at a flux limit of $\sim$ 1.0 mJy (SNR $\geq$ 5.0). 
The astrometric reference frame of FIRST maps is accurate to $\sim$ 0.05$\arcsec$ and individual sources have $\sim$ 90$\%$ confidence 
error circles of radius $<$ 0.5$\arcsec$ at the 3 mJy level and $\sim$ 1$\arcsec$ at the survey threshold.\\
In order to detect faint kpc-scale extend radio emission we also use 
the ``{\em NRAO VLA Sky Survey}'' (NVSS\footnote{http://www.cv.nrao.edu/nvss/}) which 
is a 1.4 GHz continuum survey carried out by VLA in its `D' configuration. 
NVSS covers the entire sky north of -40 deg declination and produces radio 
images with resolution of $\sim$ 45$\arcsec$ and sensitivity of $\sim$ 2.4 mJy at 5$\sigma$ level \citep{Condon98}.
NVSS observations of low resolution are advantageous in detecting low-surface-brightness KSRs that are resolved out, 
and therefore, missed by higher resolution FIRST observations.
We use the NVSS catalog which contains over 1.8 million unique detections brighter than 2.5 mJy 
and give total integrated 1.4 GHz flux densities of all the radio sources. 
The astrometric accuracy ranges from $\sim$ 1${^\prime}{^\prime}$.0 for the brightest NVSS detections to about $\sim$ 7${^\prime}{^\prime}$.0 for the 
faintest detections. 

\section{FIRST and NVSS radio counterparts}

\subsection{FIRST counterparts of Seyfert and LINER galaxies}
\begin{figure}
\includegraphics[angle=0,width=9.2cm]{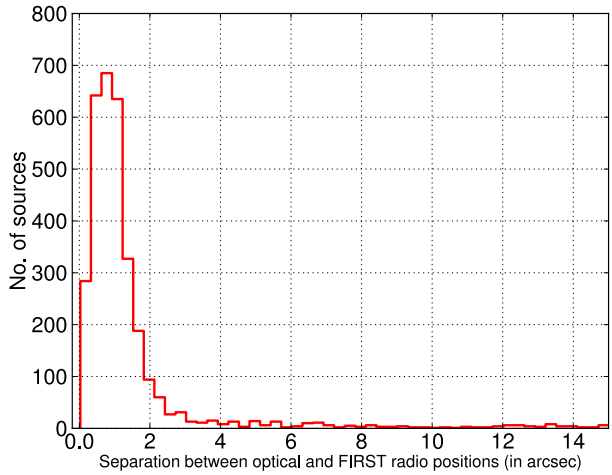}
\caption{Histogram of the separation between optical and FIRST radio positions of the cross-matched sources of LLAGN from VV10 and FIRST 
catalog.}
\label{fig:SepHistFIRST}
\end{figure}

To find the radio counterparts of Seyfert and LINER galaxies in FIRST survey we searched for FIRST detection around the optical positions 
of sources. The accurate astrometry of FIRST and optical positions of LLAGN allows us to use a simple positional matching 
with high completeness and low contamination. 
The source positions in the FIRST as well as in the VV10 catalog are measured with an accuracy better than one arcsec \citep{Becker95,Veron-Cetty10}.
The choice of matching radius between the optical and radio positions is a trade off between the completeness and contamination, 
{\ie}a smaller matching radius greatly reduces the contamination at the expense of completeness. 
In order to choose an optimum matching radius, we cross-matched our LLAGN to the FIRST sources using a large 
matching radius of 30 arcsec and plotted the histogram of the separation between the LLAGN optical position and FIRST radio position 
for all the cross-match sources. 
We find that the histogram follows nearly a Gaussian distribution between 0 to 3.0 arcsec and a flat tail onwards (see Figure~\ref{fig:SepHistFIRST}). 
Therefore, based on the histogram of the separation between LLAGN optical positions and FIRST radio positions, we choose matching radius to be 3.0 arcsec. 
Previous studies on the cross-matching between Seyfert galaxies and the FIRST sources have also reported 3.0 arcsec as a 
reliable limit with negligible probability ($<$0.05$\%$) of being merely a chance match \citep[see][]{Wadadekar04}. \\   
The cross-matching of our LLAGN sample sources and the FIRST sources using a matching radius of 3.0 arcsec results in 2651 sources.
Table~\ref{table:LLAGNNumbers} lists the number of Seyfert type 1s, Seyfert type 2s and LINER galaxies detected in FIRST. 
Before cross-matching LLAGN and FIRST sources we corrected for systematic offset between optical and radio catalogs 
({\ie}$\Delta$RA $=$ RA$_{\rm LLAGN}$ - RA$_{\rm FIRST}$ = -0.025 arcsec and $\Delta$DEC $=$ DEC$_{\rm LLAGN}$ - DEC$_{\rm FIRST}$ = -0.31 arcsec). 
Also, in cross-matching we considered only those FIRST sources that have sidelobe probability $\leq$ 0.1 ({\ie}there is 90$\%$ probability that the FIRST 
source is real).
To check the level of contamination by chance matching in our cross-matched sample we shifted the optical positions of our LLAGN by 
30 to 45 arcsec in random directions and find only 0.3$\%$ cross-matched sources. 
This shows that the contamination due to chance matching in our cross-matched sample is negligible. 
\par
To find extended radio sources with multicomponents, we searched for radio components around all the 2651 FIRST detected 
LLAGN using a large search radius of 120 arcsec. We made visual inspection of the FIRST cutout images of all the sources that show one or more 
radio components within 120 arcsec around the central core radio component. We exclude radio components that are completely unrelated to 
the source. 
This exercise resulted is a total of 180 out of 2651 LLAGN that having multicomponents detected in the FIRST survey. 
To get the total flux density of a source with multicomponents, we add up integrated flux densities of all the components.
We note that, in general, extended radio morphologies of Seyfert and LINER galaxies are more complex and diverse 
than core-jet-lobe morphology seen in radio galaxies. 
Therefore, to search for extended sources with multicomponents, it is not possible to opt for a single criterion that are often being used for quasars 
{\ie}two radio sources are located nearly symmetrically around the quasar position and the ratio of their distances to the quasar 
is `1/3 $<$ d1/d2 $<$ 3' \citep{deVries06}.
In fact, even for quasars with more complex radio morphologies ({\ie}sources with distorted asymmetric lobes and a compact core, 
or for cases in which extended lobes are resolved into complex structure), the visual inspection needs to be used \citep[see][]{Lu07}.
\subsection{NVSS counterparts of FIRST-detected Seyfert and LINER galaxies}
\begin{figure}
\includegraphics[angle=0,width=9.2cm]{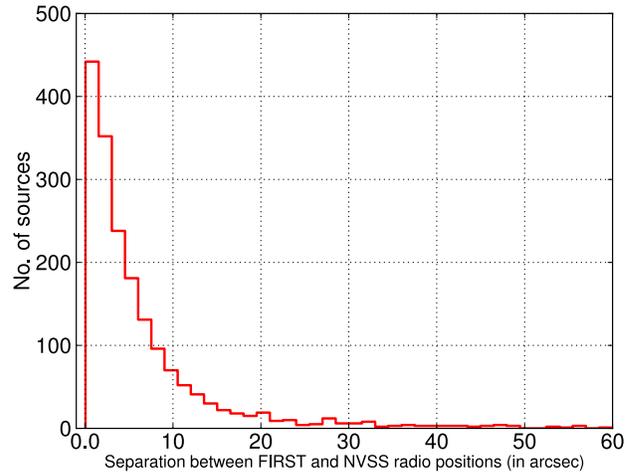}
\caption{Histogram of the separation between FIRST and NVSS radio positions.}
\label{fig:SepHistNVSS}
\end{figure}

To find the NVSS counterparts of FIRST detected LLAGN we use a search radius of 20 arcsec for all the sources that show only a single component in 
the FIRST.
The histogram of the separation between FIRST and NVSS positions of cross-matched sources tails off beyond 20 arcsec 
(see Figure~\ref{fig:SepHistNVSS}). 
Using a matching radius larger than 20 arcsec ({\eg}30 to 45 arcsec) gives only 1$\%$ to 2$\%$ increase in the number of 
cross-matched sources which is comparable to the level of chance matching. Thus, we chose optimum matching radius to 20 arcsec at which chance matching is only 0.9$\%$ 
({\ie}with FIRST source density $\sim$ 90 deg$^{-2}$). 
For extended sources with multicomponents in FIRST, we use matching radius equal to the separation between two farthest components and obtain NVSS 
counterparts of all such sources.
We made visual inspection of FIRST and NVSS image cutouts to ensure reliable cross-matching. 
The cross-matching between our FIRST detected LLAGN and NVSS catalog yields NVSS counterparts for 1737 out of 2651 sources ({\ie}65.5$\%$).
Most of FIRST detected LLAGN that remained undetected in NVSS fall below the NVSS flux limit {\ie}2.5 mJy.    
Table in the appendix lists the FIRST and NVSS parameters of our sample sources.

\section{Flux density ratio diagnostic to identify extended radio emission}

We can classify a radio source as unresolved or resolved by defining a dimensionless concentration 
parameter $\theta$ $=$  ${\rm (S_{\rm int}/S_{\rm peak})^{1/2}}$ \citep{Ivezic02}.
Using the distribution of radio sources in the two-dimensional plot of $\theta$ versus extended flux density, \cite{Kimball08} proposed that 
FIRST radio sources can be categorized as `unresolved' and `resolved' by defining $\theta$ $=$ ${\rm (S_{\rm int, FIRST}/S_{\rm peak, FIRST})^{1/2}}$ 
$\simeq$ 1.06 as a separating value. 
Therefore, we use the parameter $\theta$ $=$  ${\rm (S_{\rm int}/S_{\rm peak})^{1/2}}$ 
(square root of the ratio of integrated to peak flux density of FIRST) to infer the presence of extended radio emission at scale larger 
than the FIRST beam-size of $\sim$ 5 arcsec. 
\\
Furthermore, FIRST is a relatively high-resolution radio survey, and therefore, underestimates the flux of extended and lobe-dominated sources 
\citep{Becker95,Lu07}. Also, multiple-component radio sources with core and lobes are often detected as separate objects by FIRST 
but as only a single object in the lower-resolution NVSS observations at same wavelength. 
Therefore, the ratio between the NVSS and FIRST flux densities can be used as a diagnostic measure of 
the presence of faint extended radio emission that is missed by FIRST. 
\cite{Kimball08} showed that the distribution of the difference of FIRST and NVSS radio magnitudes $\Delta$t $=$ t$_{\rm FIRST}$ - t$_{\rm NVSS}$ is 
bimodal with two peaks centered at $\Delta$t $=$ 0 and $\Delta$t $=$ 0.7; 
where radio magnitude is defined as t $=$ -2.5 log($\frac{\rm S_{\rm int}}{\rm 3631~Jy}$). 
The visual inspection of thousand of FIRST images confirmed that 
sources at the $\Delta$t $=$ 0 locus are single component sources, while those in the $\Delta$t $=$ 0.7 locus are multiple-component or extended. 
By fitting the bimodal distribution of  $\Delta$t with two Gaussians \cite{Kimball08} proposed that 
$\Delta$t $=$ t$_{\rm FIRST}$ - t$_{\rm NVSS}$ $\simeq$ 0.35 can be used as a separating value between `simple' ($\Delta$t $<$ 0.35) and `complex' 
($\Delta$t $>$ 0.35) radio morphologies. 
In terms of flux density ratio, the difference in FIRST and NVSS radio magnitude $\Delta$t = 0.35 corresponds 
to $\frac{\rm S_{\rm int, NVSS}}{\rm S_{\rm int, FIRST}}$ $\simeq$ 1.38.  
Hence, we define a parameter ${\theta}_{\rm NVSS-FIRST}$ $=$  ${\rm (S_{\rm NVSS, int}/S_{\rm FIRST, int})^{1/2}}$ with separating value 
${\theta}_{\rm NVSS-FIRST}$ $=$ 1.175 ({\ie}corresponding $\Delta$t = 0.35) 
to classify radio sources into `simple' (${\theta}_{\rm NVSS-FIRST}$ $\leq$ 1.175) 
and `complex' (${\theta}_{\rm NVSS-FIRST}$ $>$ 1.175) radio morphology categories.  
\par
Therefore, using FIRST and NVSS flux densities we can divide radio sources into following classes.\\
(i) `Unresolved' point sources in FIRST with `simple' radio morphology inferred from FIRST-NVSS flux density comparison. 
These sources can be characterized with ${\theta}_{\rm FIRST}$ $\leq$ 1.06 and ${\theta}_{\rm NVSS-FIRST}$ $\leq$ 1.175.\\
(ii) `Unresolved' sources in FIRST but with `complex' radio morphology inferred from FIRST-NVSS flux density comparison. 
These sources can be characterized with ${\theta}_{\rm FIRST}$ $\leq$ 1.06 and ${\theta}_{\rm NVSS-FIRST}$ $>$ 1.175. 
These sources possess extended low-surface-brightness radio emission that is resolved out in FIRST but detected in NVSS 
due to larger beam-size. \\
(iii) `Resolved' sources in FIRST but `simple' radio morphology inferred from FIRST-NVSS flux density comparison. 
These sources can be characterized with ${\theta}_{\rm FIRST}$ $>$ 1.06 and ${\theta}_{\rm NVSS-FIRST}$ $\leq$ 1.175. 
These sources do have extended radio emission detected in FIRST but there is no significant additional flux detected by NVSS.  
\\
(iv) `Resolved' sources in FIRST and `complex' radio morphology inferred from FIRST-NVSS comparison. 
These sources can be characterized with ${\theta}_{\rm FIRST}$ $>$ 1.06 and ${\theta}_{\rm NVSS-FIRST}$ $>$ 1.175. 
These sources do have extended radio emission detected in FIRST and also have additional faint low-surface-brightness radio emission component 
detected in NVSS.
\\    
In interpreting the results based on the flux density ratios we should keep the caveat in mind that FIRST has higher sensitivity than NVSS.  
Table~\ref{table:RadioMorph} lists the number and fraction of Seyferts/LINER galaxies with unresolved/resolved 
and simple/complex radio morphologies.

\begin{table*}
\centering
\begin{minipage}{140mm}
\caption{Radio morphology}
\begin{tabular}{@{}ccccccc@{}}
\hline
Radio morphology   & ${\theta}_{\rm FIRST}$ & ${\theta}_{\rm NVSS-FIRST}$ & \multicolumn{4}{c}{No. of sources}  \\ 
                   &                       &                              &  All    &    S1    & S2    & S3     \\ \hline
{\it FIRST detected sources}               &                 &               & 2651  &  1373    &  1164  &  114  \\
                                           &                 &               &       &          &       &         \\
     Unresolved                            &  $\leq$ 1.06    &               & 1824 (68.8$\%$) & 913 (66.5$\%$) & 839 (72.1$\%$)& 72 (63.2$\%$) \\
     Resolved                              &  $>$ 1.06       &               & 827 (31.2$\%$)  & 460 (33.5$\%$) & 325 (27.9$\%$)& 42 (36.8$\%$)  \\
                                           &                 &               &       &          &       &         \\
 {\it FIRST - NVSS pair}                   &                 &               & 1737  &  884     &  753  &  100    \\
                                           &                 &               &       &          &       &         \\
Unresolved and simple             & $\leq$ 1.06  & $\leq$ 1.175    & 961 (55.3$\%$) & 463 (52.4$\%$) & 447 (59.4$\%$) & 51 (51$\%$) \\
Unresolved and complex            & $\leq$ 1.06  & $>$ 1.175       & 196 (11.3$\%$) & 94 (10.6$\%$) & 90 (11.9$\%$) & 12 (12$\%$)    \\ 
Resolved and simple               &  $>$ 1.06     & $\leq$ 1.175     & 483 (27.8$\%$) &  272 (30.8$\%$)  &  181 (24$\%$)  &  30 (30$\%$)      \\
Resolved and complex              &    $>$ 1.06    &   $>$ 1.175     & 97 (5.6$\%$) &  55 (6.2$\%$) & 35 (4.6$\%$) & 7 (7$\%$)   \\ \hline
\end{tabular}
\label{table:RadioMorph} 
\\ \\
Note : Unresolved and resolved morphology is based on the ratio of total to peak FIRST flux densities (${\theta}_{\rm FIRST}$), 
while simple and complex radio morphologies are based on the ratio of NVSS to FIRST total flux densities (${\theta}_{\rm NVSS-FIRST}$).
\end{minipage}
\end{table*}

\subsection{Comparison of total to peak flux densities in FIRST}

\begin{figure*}
\includegraphics[angle=0,width=9.2cm]{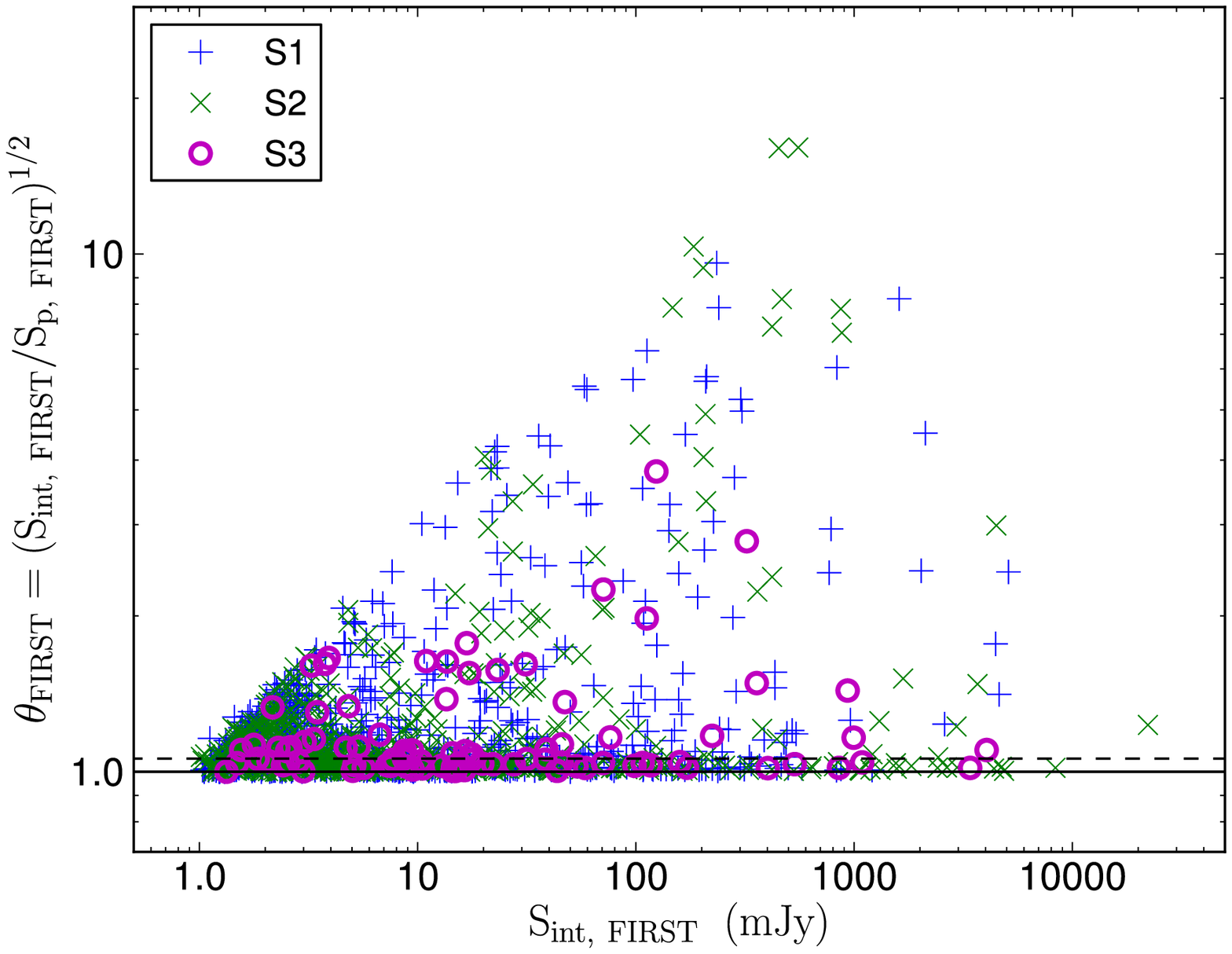}{\includegraphics[angle=0,width=9.2cm]{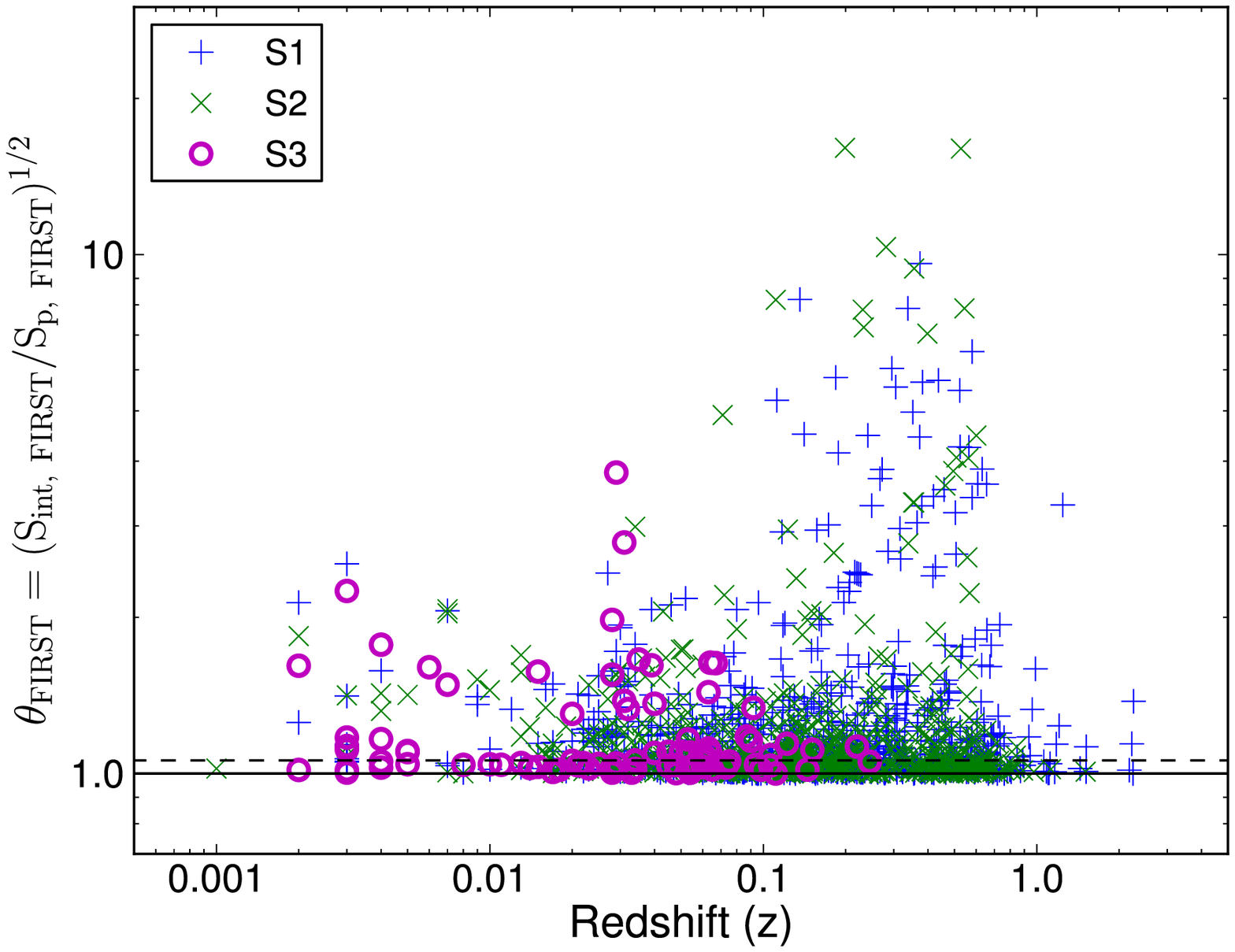}}
\caption{{\it Left :} ${\theta}_{\rm FIRST}$ versus S$_{\rm int, FIRST}$. {\it Right :} ${\theta}_{\rm FIRST}$ versus Redshift. Seyfert 1s, Seyfert 2s and LINERs are shown by `+', `$\times$' and circles, 
respectively. Solid and dashed horizontal lines represents ${\theta}_{\rm FIRST}$ $=$ 1.0 and ${\theta}_{\rm FIRST}$ $=$ 1.06, respectively. 
Unresolved sources (${\theta}_{\rm FIRST}$ $\leq$ 1.06) are lying in between solid and dashed lines, while resolved sources 
(${\theta}_{\rm FIRST}$ $>$ 1.06) are lying above the dashed line.}
\label{fig:ThetaVsFluxRedshift}
\end{figure*}

Figure~\ref{fig:ThetaVsFluxRedshift} (left panel) shows the distribution of FIRST total flux density (S$_{\rm int, FIRST}$) versus 
${\theta}_{\rm FIRST}$ $=$  ${\rm (S_{\rm int, FIRST}/S_{\rm peak, FIRST})^{1/2}}$. 
It is evident that resolved sources with extended radio emission larger than 5$\arcsec$ ({\ie}${\theta}_{\rm FIRST}$ $>$ 1.06) 
are present across all flux densities. 
Although, sources with high ratios of total$-$to$-$peak flux densities are found only in radio bright LLAGN. 
For example, sources with ${\theta}_{\rm FIRST}$ $>$ 2.0 ({\ie}S$_{\rm int}$ higher than 4 times of S$_{\rm peak}$) are found only 
at S$_{\rm int}$ $\geq$ 5.0 mJy. 
The comparison of FIRST total to peak density (${\theta}_{\rm FIRST}$) shows that 827/2615 $\sim$~31.2$\%$ sources possess extended radio emission on 
scales larger than 5 arcsec (see Table~\ref{table:RadioMorph}). 
We note that the resolved sources with extended radio emission (${\theta}_{\rm FIRST}$ $>$ 1.06) are distributed across all redshifts spanning 
over 0.002 to 2.26 with median redshift $\sim$ 0.136 (figure~\ref{fig:ThetaVsFluxRedshift}, right panel). 
There are 794/2651 $\sim$ 30$\%$ sources with redshift ($z$) $\geq$ 0.01, where 5 arcsec angular size corresponds to $\geq$ 1.0 kpc. 
Therefore, we infer that 30$\%$ FIRST detected LLAGN possess KSRs {\ie}radio emitting structures larger than 1.0 kpc. 
This is only a lower limit as sources with radio structures of $\geq$ 1.0 kpc would appear unresolved at higher redshifts ($z$ $\geq$ 0.01).
Also, LLAGN with AGN-jet lying close to the line-of-sight (in type 1s) would have much smaller projected size and therefore, would 
appear as unresolved despite possessing extended radio emission. 
\subsection{Comparison of NVSS and FIRST flux densities}

\begin{figure*}
\includegraphics[angle=0,width=9.2cm]{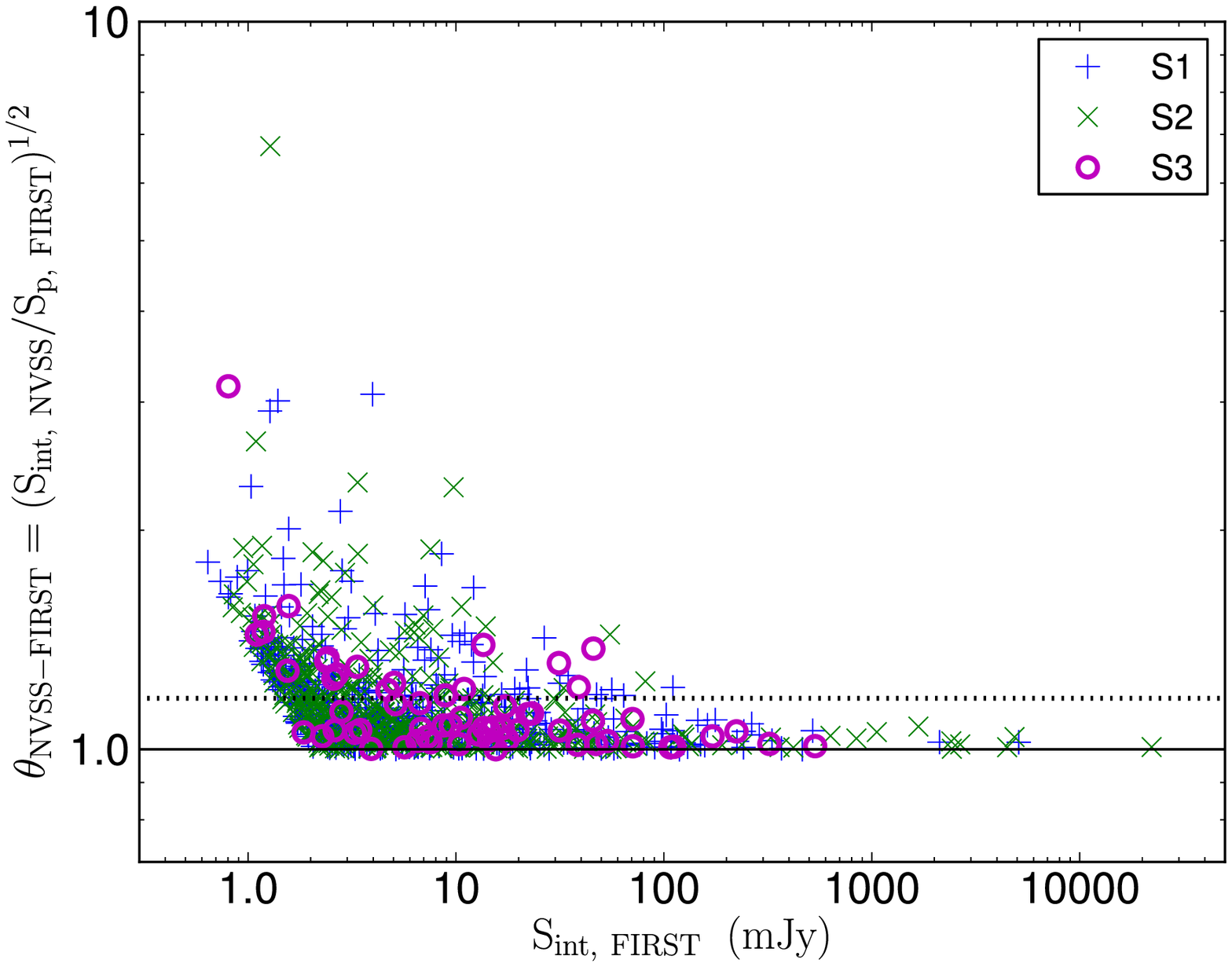}{\includegraphics[angle=0,width=9.2cm]{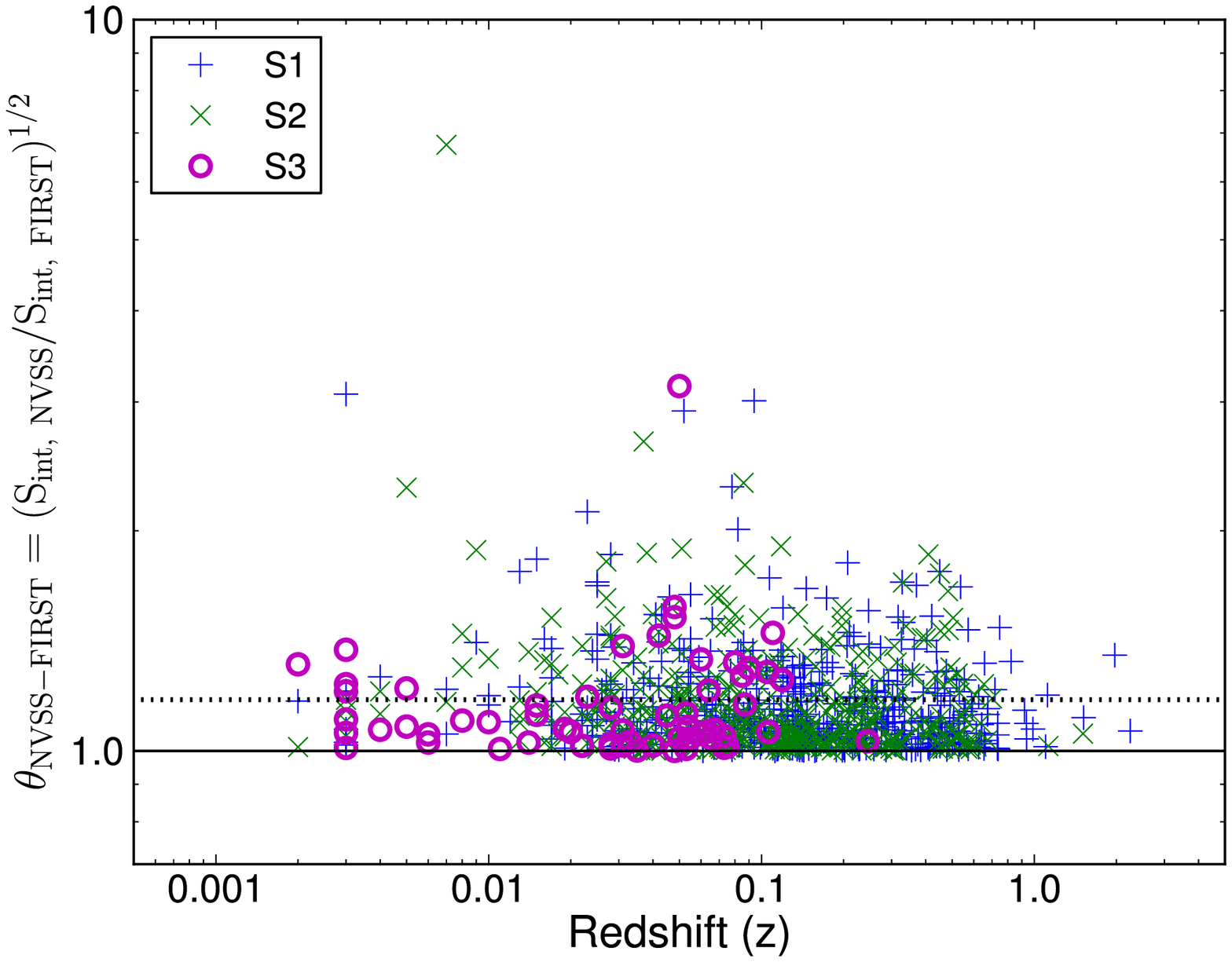}}
\caption{{\it Left :} ${\theta}_{\rm NVSS-FIRST}$ versus S$_{\rm int, FIRST}$. {\it Right :} ${\theta}_{\rm NVSS-FIRST}$ versus redshift. Seyfert 1s, Seyfert 2s and LINERs are shown by `+', `$\times$' and circles, 
respectively. Solid and dotted horizontal lines represents ${\theta}_{\rm NVSS-FIRST}$ $=$ 1.0 and ${\theta}_{\rm NVSS-FIRST}$ $=$ 1.175, respectively. 
Simple sources (${\theta}_{\rm NVSS-FIRST}$ $\leq$ 1.175) are lying in between solid and dotted lines, while complex sources 
(${\theta}_{\rm NVSS-FIRST}$ $>$ 1.175) are lying above the dotted line.}
\label{fig:ThetaNVSSVsFluxRedshift}
\end{figure*}

The faint extended low-surface-brightness radio emission that is resolved out in FIRST observations can be detected in NVSS 
due to its much larger 45$\arcsec$ beam. 
Therefore, we also compare NVSS and FIRST total flux densities by 
defining a parameter ${\theta}_{\rm NVSS-FIRST}$ $=$ ${\rm (S_{\rm int, NVSS}/S_{\rm int, FIRST})^{1/2}}$ which is used to infer 
the radio morphologies of our sample sources. 
Figure~\ref{fig:ThetaNVSSVsFluxRedshift} (Left panel) shows the distribution 
of FIRST total flux density (S$_{\rm int, FIRST}$) versus ${\theta}_{\rm NVSS-FIRST}$ $=$ ${\rm (S_{\rm int, NVSS}/S_{\rm int, FIRST})^{1/2}}$. 
It is clear that sources characterized with complex morphology ({\ie}${\theta}_{\rm NVSS-FIRST}$ $>$ 1.75) 
are present in radio faint as well as in radio bright LLAGN. 
However, there is no additional NVSS flux density detected in very bright radio sources {\ie}S$_{\rm int, FIRST}$ $\geq$ 100 mJy. 
Table~\ref{table:RadioMorph} lists the number and fraction of LLAGN that possess additional flux density detected in NVSS but missed out in FIRST. 
We note that the extra flux detected in NVSS is present in resolved as well as unresolved FIRST sources. 
The comparison of NVSS and FIRST flux densities reveals the presence of complex radio morphology ({\ie} ${\theta}_{\rm NVSS-FIRST}$ $>$ 1.75 ) 
in $\sim$ 17$\%$ of FIRST-NVSS detected sources.  
Interestingly, there are $\sim$ 11$\%$ FIRST-NVSS detected LLAGN in which FIRST detects only unresolved compact emission 
(${\theta}_{\rm FIRST}$ $<$ 1.06) while NVSS picks up additional flux density (${\theta}_{\rm FIRST}$ $>$ 1.175). 
Thus, in $\sim$ 11$\%$ of sources, the comparison of NVSS and FIRST flux densities helps to recover low-surface-brightness extended 
radio emission that is missed in FIRST observations. 
Total fraction of sources with extended radio emission identified either via FIRST (${\theta}_{\rm FIRST}$) or via 
NVSS (${\theta}_{\rm NVSS-FIRST}$) flux diagnostics is as high as $\sim$ 45$\%$ among the FIRST-NVSS detected LLAGN 
(see Table~\ref{table:RadioMorph}). 
The ${\theta}_{\rm NVSS-FIRST}$ versus redshift ($z$) plot (see Figure~\ref{fig:ThetaNVSSVsFluxRedshift}; right panel) shows 
that complex sources are present across all redshifts. 
Among the FIRST-NVSS detected LLAGN there are 911/1737 $\sim$ 42$\%$ resolved/complex sources lying at 
redshift $z$ $\geq$ 0.01 and thus possess KSRs ({\ie}5$\arcsec$ angular scale corresponds to 1.0 kpc at $z$ $\geq$ 0.01).   
We note that this fraction is only a lower limit as it is possible that in sources of bright compact component with very faint 
extended emission, the total flux is primarily due to the brighter compact component, thus, resulting 
${\theta}_{\rm NVSS-FIRST}$ $=$ ${\rm (S_{\rm int, NVSS}/S_{\rm int, FIRST})^{1/2}}$ $\leq$ 1.06 with source being classified as compact. 
\par
\begin{figure}
\includegraphics[angle=0,width=9.2cm]{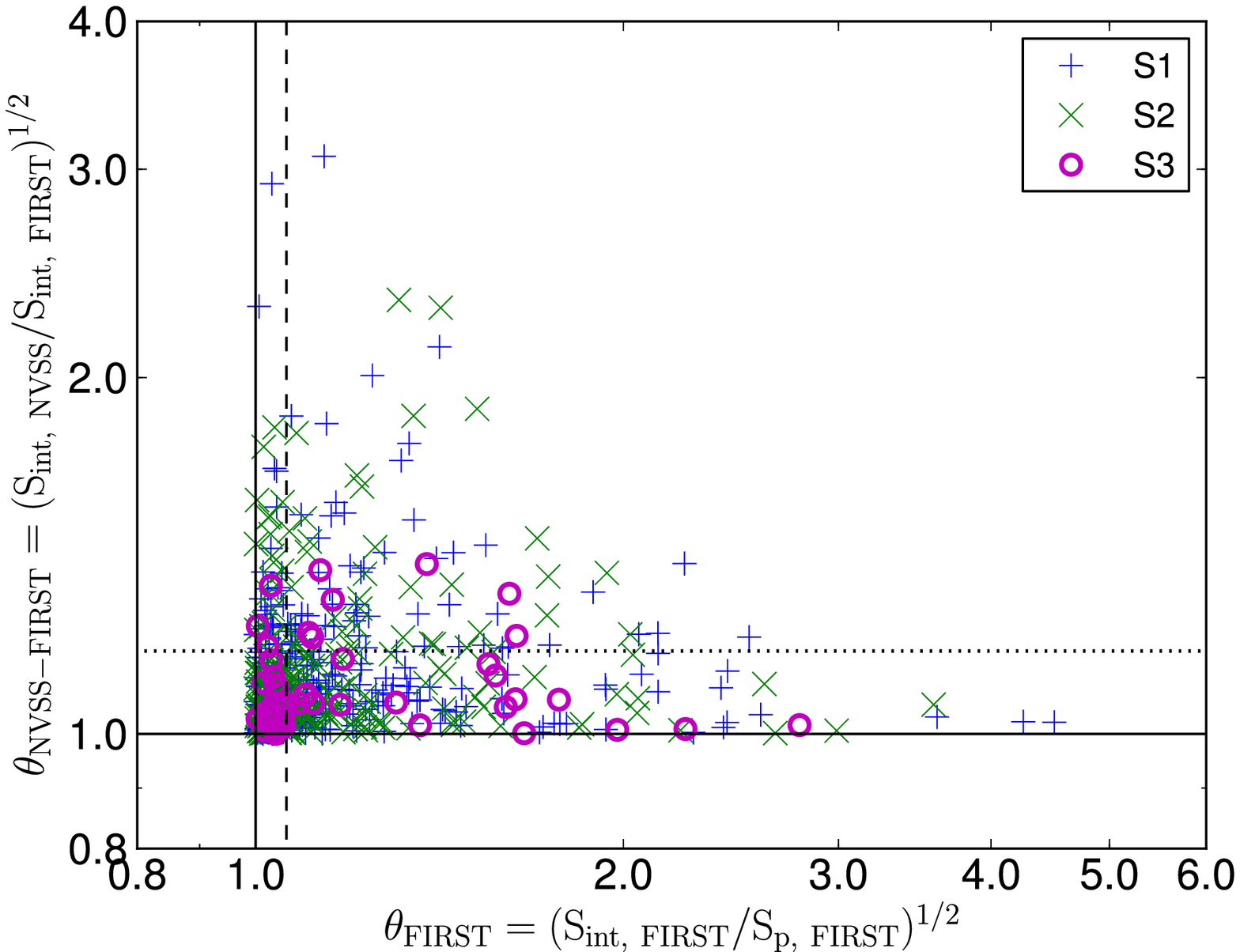}
\caption{${\theta}_{\rm NVSS-FIRST}$ versus ${\theta}_{\rm FIRST}$. Seyfert 1s, Seyfert 2s and LINERs are shown by `+', `$\times$' and circles, 
respectively. Dashed vertical line represents dividing line (${\theta}_{\rm FIRST}$ $=$ 1.06) between resolved and unresolved sources 
detected in FIRST. 
Dotted horizontal line represents dividing line (${\theta}_{\rm NVSS-FIRST}$ $=$ 1.175) between simple and complex sources detected in NVSS. 
It is evident that there is no systematic trend and sources of variety of radio morphologies are present in our sample.}
\label{fig:ThetaFIRSTVsThetaNVSS}
\end{figure}

Figure~\ref{fig:ThetaFIRSTVsThetaNVSS} shows ${\theta}_{\rm FIRST}$ versus ${\theta}_{\rm NVSS-FIRST}$ plot. 
This plot does not show any systematic trend suggesting that there are wide variety of radio sources 
{\ie}(i) unresolved sources in both FIRST and NVSS, (ii) unresolved sources in FIRST but extended emission detected in NVSS, 
(iii) resolved sources in FIRST with no extra flux density in NVSS, and (iv) resolved sources in FIRST with additional flux density detected in NVSS. 
Indeed, this is also evident from table~\ref{table:RadioMorph} that lists the fraction of sources with different radio morphologies.
\par
Table~\ref{table:KSRFraction} shows the comparison of the fraction of KSR sources in our sample and previous 
studies. It evident that high-resolution radio observations carried out at higher frequencies miss the detection of KSRs. 
While low-frequency observations with relatively lower resolution can efficiently detect kpc-scale radio emission of low-surface-brightness.    
Noting the importance of low-frequency observations with relatively low-resolution in detecting KSRs, 
we have carried out 325 MHz and 610 MHz Giant Metrewave Radio Telescope (GMRT) observations of a 
carefully chosen sample of Seyfert galaxies \cite[{\eg}][]{Kharb14}. 
Our investigation on faint KSRs using low-frequency GMRT observations is expected to conclude soon (Singh et al. 2015, in progress).   
\section{LLAGN with multicomponent radio emission detected in FIRST}
\begin{figure*}
\includegraphics[angle=0,width=5.8cm,trim={0 0 1.5cm 0},clip]{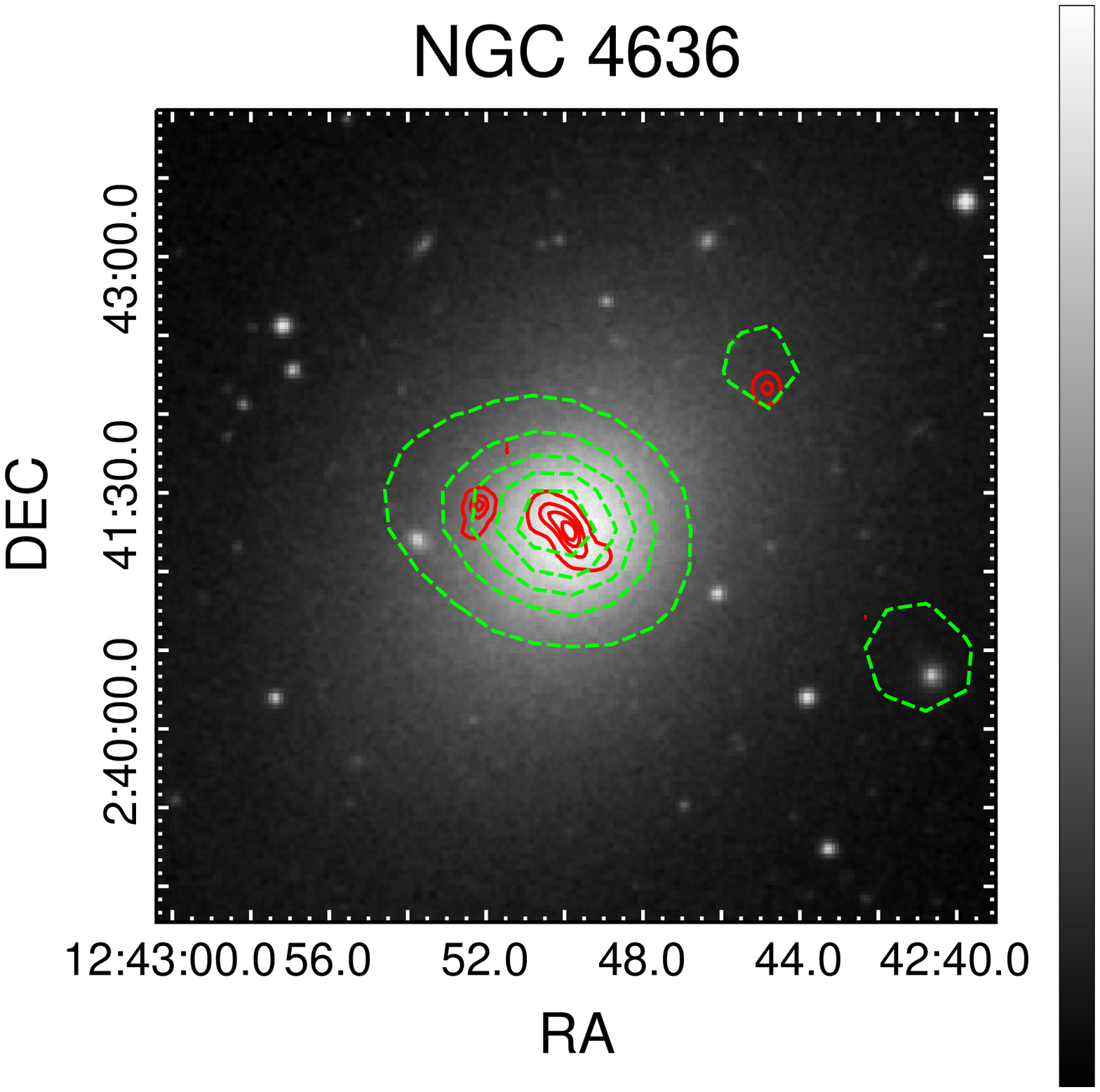}{\includegraphics[angle=0,width=5.8cm,trim={0 0 1.5cm 0},clip]{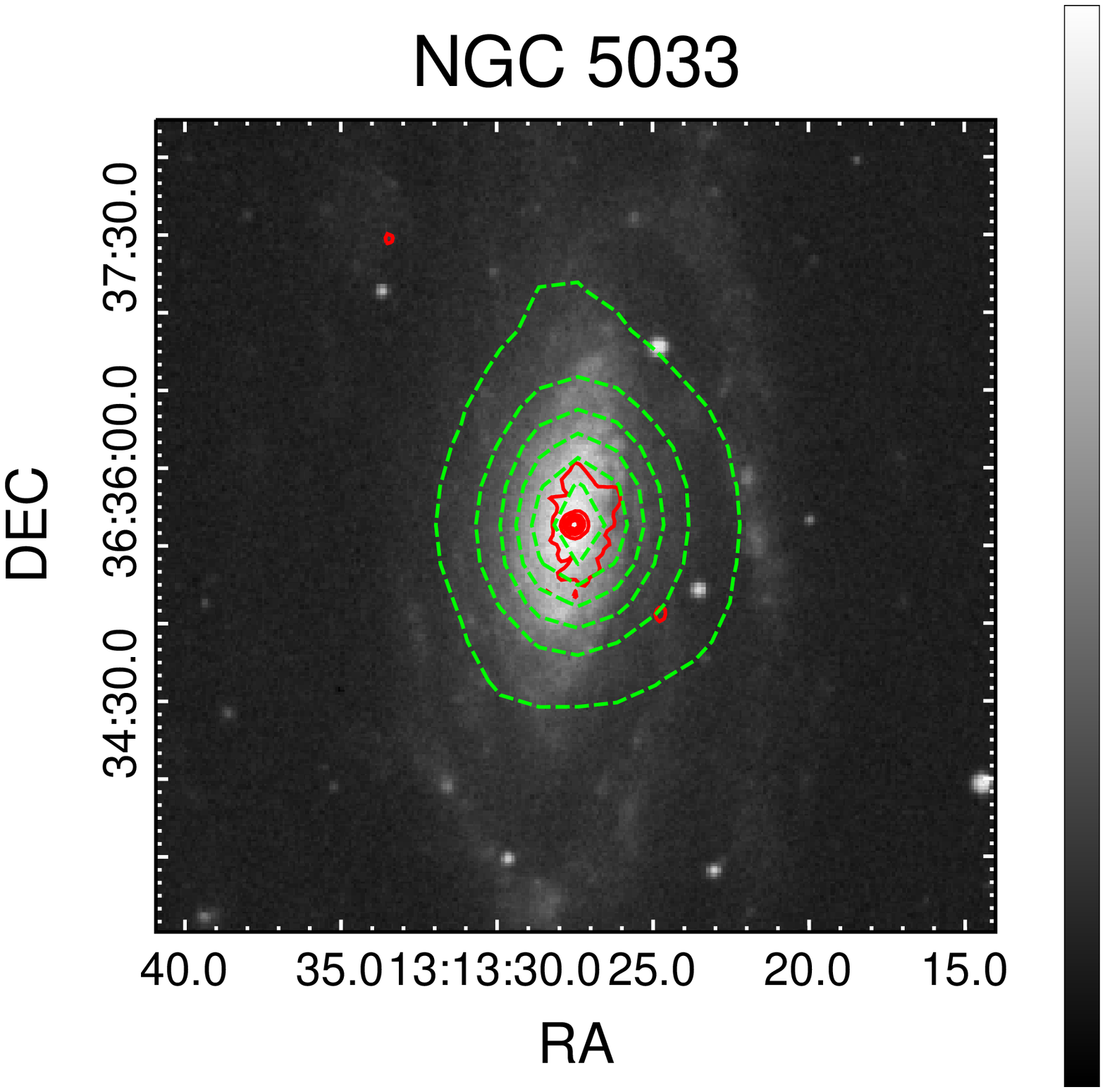}}
{\includegraphics[angle=0,width=5.8cm,trim={0 0 1.5cm 0},clip]{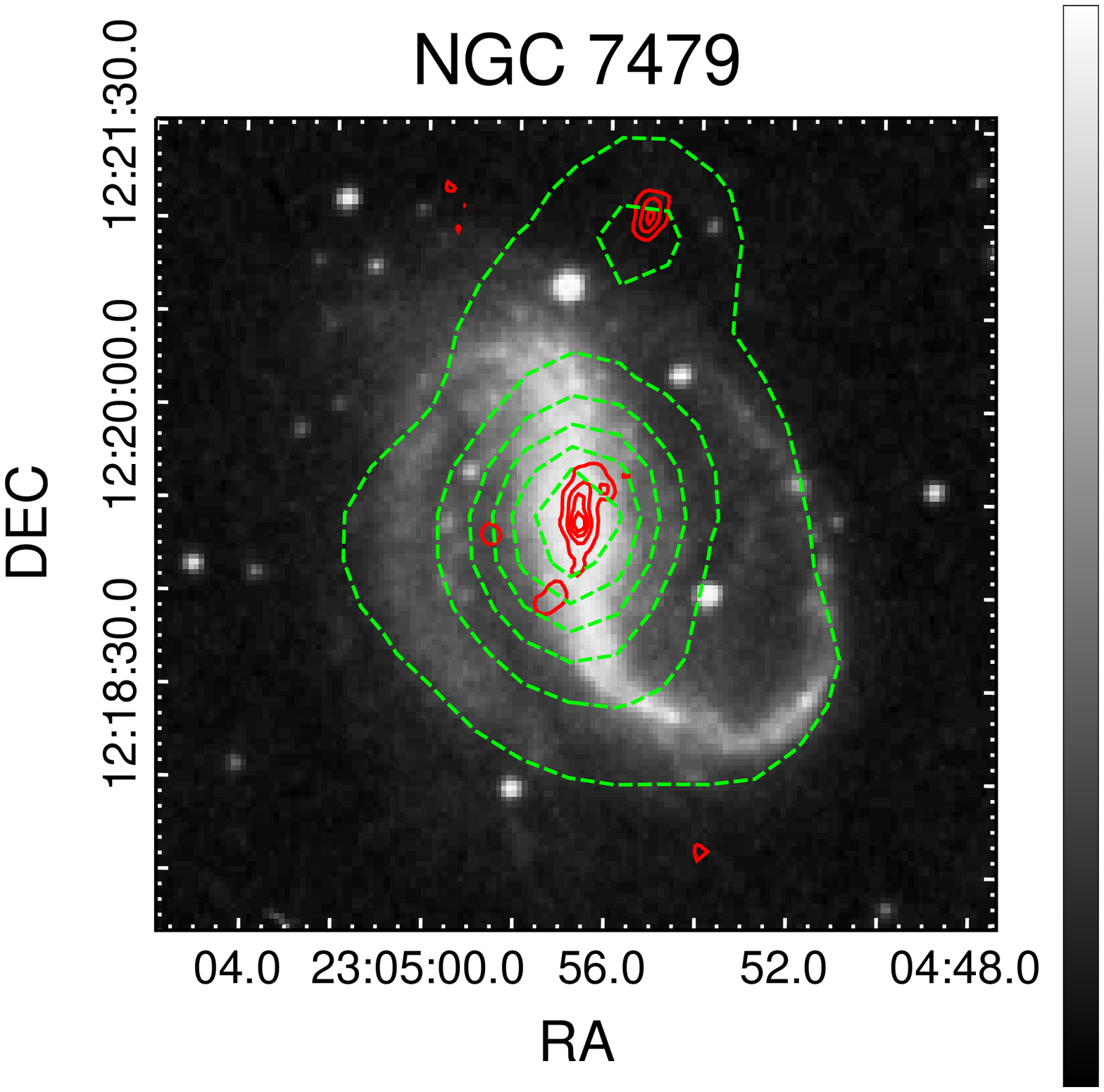}}
\caption{FIRST contours (in red solid curves) and NVSS contours (in green dotted curves) overplotted on the grey scale DSS optical images for 
NGC 4636 (left panel), NGC 5033 (middle panel) and NGC 7479 (right panel).} 
\label{fig:RadioCont}
\end{figure*}

There are a total of 180 Seyfert and LINER galaxies in our sample that display multicomponent radio emission in FIRST observations.  
These sources have more than one components wherein each component is fitted with an elliptical Gaussian.
In figure~\ref{fig:RadioCont}, we show FIRST and NVSS contours overplotted on the DSS optical images for three sources 
({\ie}NGC 4636, NGC 5033 and NGC 7479) that display extended multicomponent radio emission. \\
(i) 
FIRST observations of NGC 4636, a LINER galaxy at redshift ($z$) $\sim$ 0.003, show an S-shaped radio structure with projected size 
of $\sim$ 2.48 kpc ($\sim$ 40 arcsec) along Position Angle (PA) $\sim$ 40$^{\circ}$ (see Figure~\ref{fig:RadioCont}, left panel). 
There is another radio component located at the distance of $\sim$ 34$^{\arcsec}$ to the west of the central component. 
The jet-like radio structure accompanied by a strong ridge of radio emission displaying an S-shaped structure of 
$\sim$ 5.0 kpc has been reported in previous observations \citep[see][]{Birkinshaw85}. 
The linear jet-like radio structure lies nearly along the minor axis of host galaxy \citep{Stanger86}.\\  
(ii) The FIRST image of NGC 5033, a Seyfert type 1.8 galaxy at redshift ($z$) $\sim$ 0.003, shows a central radio component accompanied by a diffuse extended 
radio structures with projected size of $\sim$ 3.6 kpc ($\sim$ 60$^{\arcsec}$) along PA $\sim$ 135$^{\circ}$ 
(see Figure~\ref{fig:RadioCont}, middle panel). 
The diffuse extended radio emission appear to align with the host galaxy major axis. 
This source possesses a complex radio structure with a slightly resolved core surrounded by a diffuse envelope of radio structure 
of $\sim$ 10$^{\arcsec}$ (0.9 kpc), roughly along the east-west direction \citep{Nagar02}. 
The tapered maps have shown the ridge of emission spanning over $\sim$ 3.6 kpc in north-south direction 
along the galaxy major axis. \\
(iii) 
NGC 7479, a Seyfert 1.9 galaxy at redshift ($z$) 0.0079, exhibits a curve-shaped extended emission along the north direction and a weaker southern 
component. The total extension of north to south component is $\sim$ 9.2 kpc (56$^{\arcsec}$).    
\cite{Beck02} reported that NGC 7479 has strong polarized radio emission, mainly due to the nuclear jet. \\
The detailed study of the radio structures in all our sample LLAGN with multicomponent radio emission will be presented in our 
next paper. The FIRST images of three sources presented here show that KSRs in 
Seyferts and LINERs exhibit varying morphologies {\eg}linear, S-shaped, diffuse. 
This is consistent with previous studies based on more sensitive targeted radio observations of small Seyfert samples 
\citep[{\eg}][]{Ulvestad89,Kukula93,Baum93,Colbert96,Gallimore06}.

\begin{table*}
\centering
\begin{minipage}{140mm}
\caption{Comparison of detected fraction KSR sources}
\begin{tabular}{@{}cccccccc@{}}
\hline
Reference    & $\nu$ & Telescope & Resolution & Sensitivity (5$\sigma$) &  N(Total) & N(KSRs) & Detection fraction \\ 
          &  (GHz)  & Configuration & (arcsec)   &  (mJy)       &            &         &            \\ \hline
\cite{Thean2000}   &  8.4  &  VLA A   &  $\sim$ 0.25   & $\sim$ 0.28   & 60   & 4   & $\sim$ 6.7$\%$    \\
\cite{Kukula95}    &  8.4  &  VLA A   &  $\sim$ 0.25   & $\sim$ 0.35   & 18   & 0   & 0.0 $\%$     \\
\cite{Ulvestad89}  &  1.4  &  VLA A/B &  $\sim$ 2.0    & $\sim$ 0.40   & 57   & 13  & $\sim$ 23$\%$    \\
\cite{Gallimore06} &  5.0  &  VLA D   &  $\sim$ 15 - 20& $\sim$ 0.25   & 43   & 19  & $\sim$ 44$\%$     \\
{\it Our work}           &       &          &                &               &      &     &            \\ 
FIRST sample       &  1.4  & VLA B    &  $\sim$ 5      & $\sim$ 1.0    & 2651 & 794 & $\sim$ 30.0$\%$   \\ 
FIRST-NVSS sample  &  1.4  & VLA D    &  $\sim$ 45     & $\sim$ 2.5    & 1737 & 738 & $\sim$ 42.3$\%$   \\ \hline
\end{tabular}
\label{table:KSRFraction} 
\end{minipage}
\end{table*}

\section{Nature of kpc-scale radio emission} 
We attempt to investigate the nature and origin of KSRs detected in Seyfert and LINER galaxies of our sample. 
We study radio luminosity distributions, radio-loudness parameters, the dependence KSRs radio power 
to the AGN power, ratios of radio$-$to$-$FIR fluxes and mid-IR colors to derive the statistical inferences about the origin of KSRs. 
\subsection{Radio luminosities}
\begin{figure}
\includegraphics[angle=0,width=9.2cm]{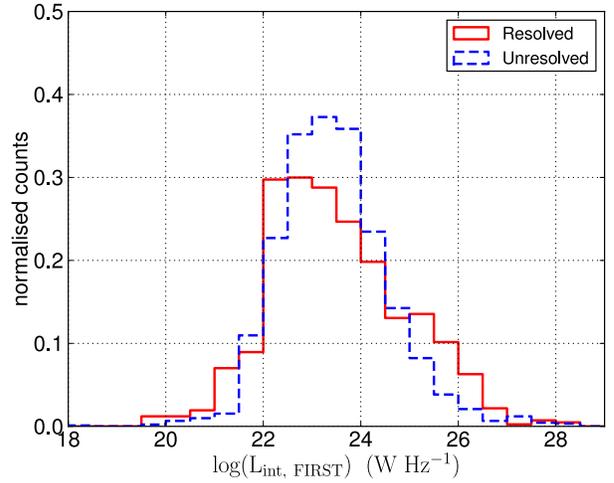}
\caption{Distributions of FIRST 1.4 GHz radio luminosities (L$_{\rm int, FIRST}$) for unresolved and resolved sources.}
\label{fig:LIntFirst}
\end{figure}

Figure~\ref{fig:LIntFirst} shows the radio luminosity distributions of unresolved and resolved sample sources detected in FIRST. 
It is evident that the luminosity distribution for unresolved sources can be represented by a normal distribution, 
while radio luminosity distribution of resolved sources is skewed towards higher radio luminosity. 
Although, both unresolved sources (${\theta}_{\rm FIRST}$ $\leq$ 1.06) and resolved sources (${\theta}_{\rm FIRST}$ $>$ 1.06) have 
similar median radio luminosity values $\sim$ 2.25 $\times$ 10$^{23}$ W Hz$^{-1}$ and $\sim$ 1.98 $\times$ 10$^{23}$ W Hz$^{-1}$, respectively. 
The skewed distribution of resolved sources indicates that the resolved source population is relatively more dominant 
at higher radio luminosities.  
\par
\begin{figure*}
\includegraphics[angle=0,width=9.2cm]{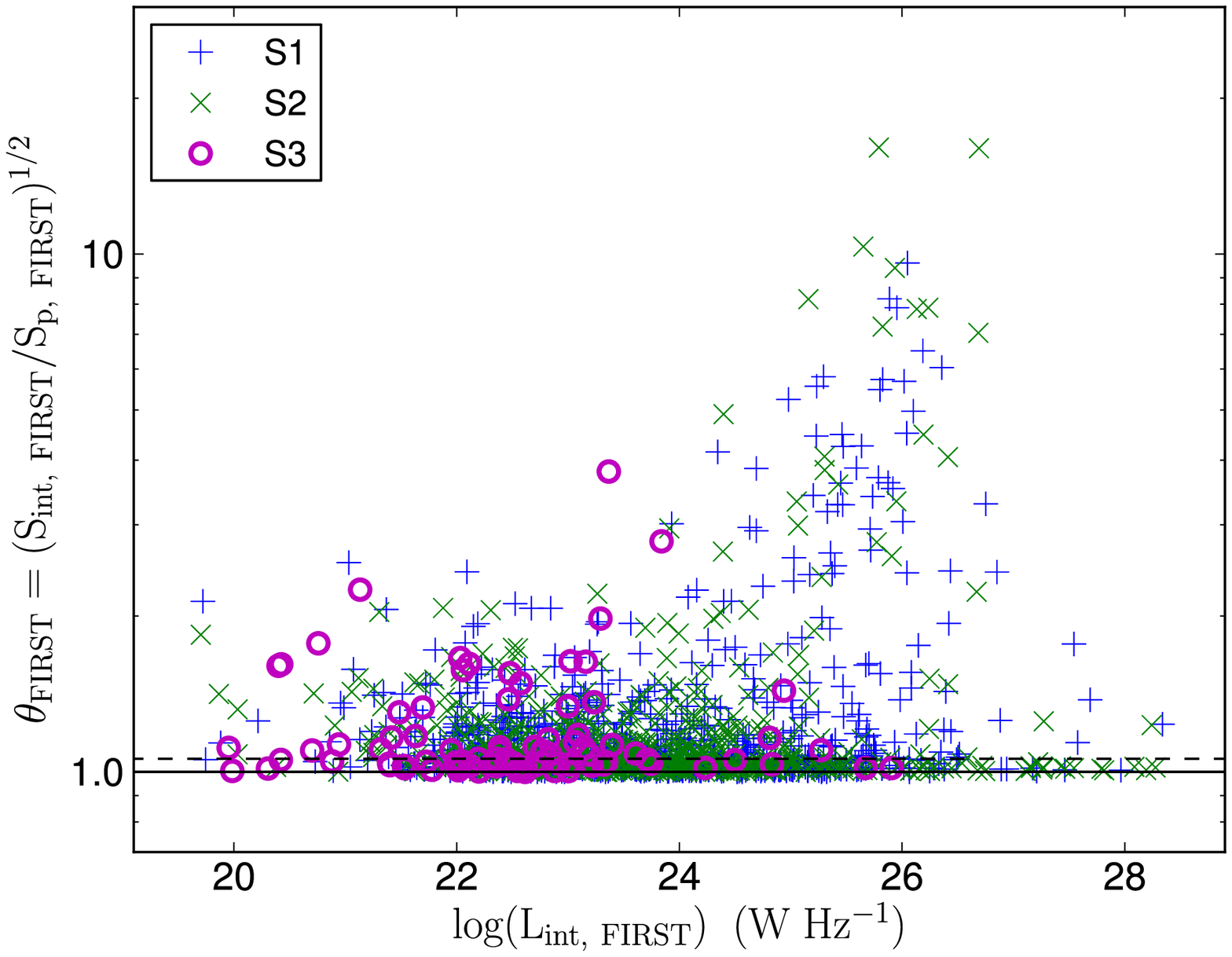}{\includegraphics[angle=0,width=9.2cm]{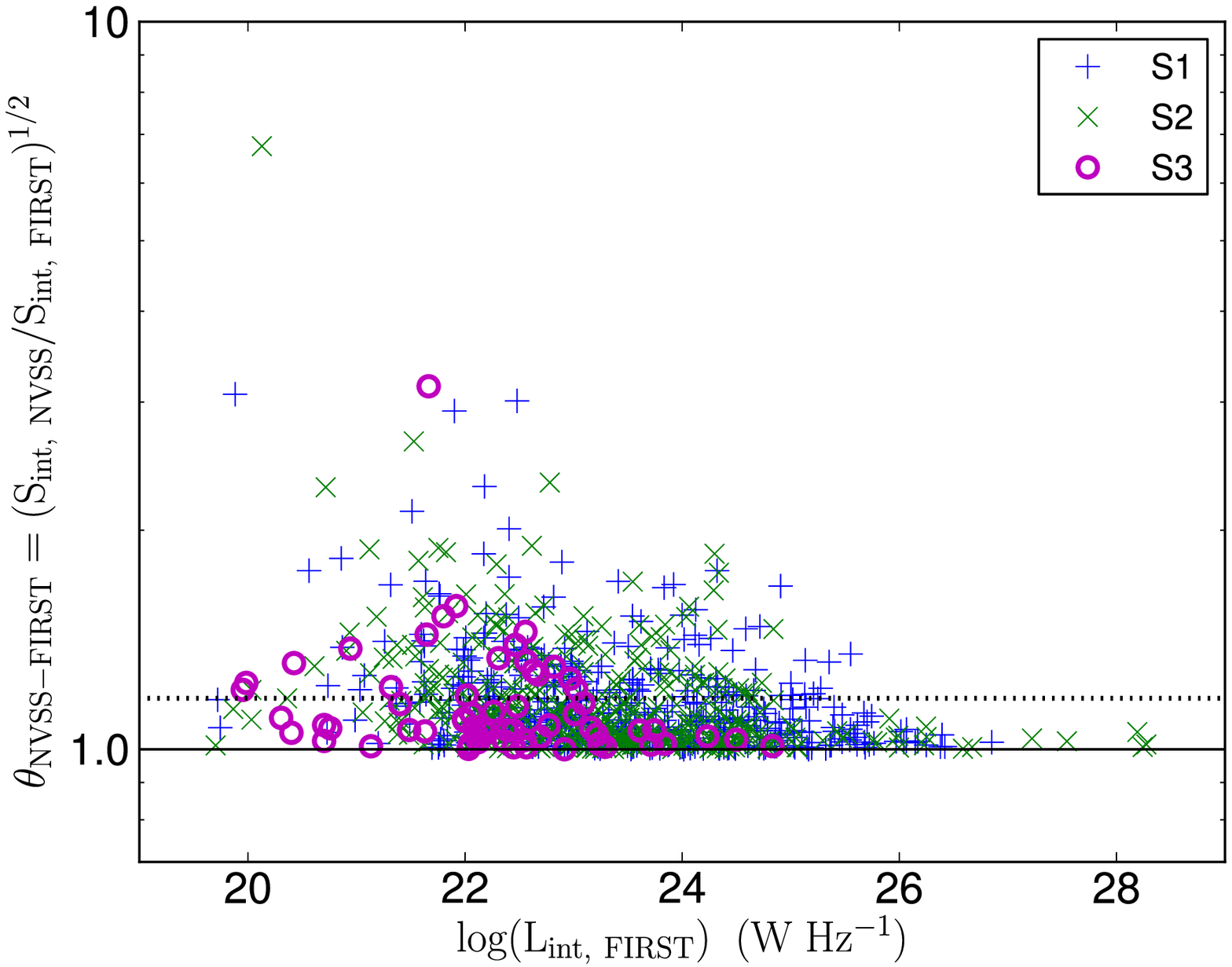}}
\caption{{\it Left}: ${\theta}_{\rm FIRST}$ versus L$_{\rm 1.4~GHz,~FIRST}$. 
Dashed horizontal line represents (${\theta}_{\rm FIRST}$ $=$ 1.06) dividing line between unresolved and resolved sources. 
{\it Right}:  ${\theta}_{\rm NVSS-FIRST}$ versus L$_{\rm 1.4~GHz,~FIRST}$. 
Dotted horizontal line represents dividing line (${\theta}_{\rm NVSS-FIRST}$ $=$ 1.175) between simple and complex sources detected in NVSS. 
Seyfert 1s, Seyfert 2s and LINERs are shown by `+', `$\times$' and circles, respectively.}
\label{fig:LFIRSTVsThetaFIRST}
\end{figure*}

To examine if the occurrence of KSRs is related to the total radio luminosity of a source 
we plot extended emission parameters ${\theta}_{\rm FIRST}$ and ${\theta}_{\rm NVSS-FIRST}$ versus 1.4 GHz FIRST total radio luminosity (L$_{\rm int,~FIRST}$). 
Figure~\ref{fig:LFIRSTVsThetaFIRST} shows the plots of ${\theta}_{\rm FIRST}$ versus L$_{\rm int,~FIRST}$ (left panel), 
and ${\theta}_{\rm NVSS-FIRST}$ versus L$_{\rm int,~FIRST}$ (right panel). 
We note that resolved and complex sources ({\ie}sources with ${\theta}_{\rm FIRST}$ $>$ 1.06 and ${\theta}_{\rm NVSS-FIRST}$ $>$ 1.175) are 
present across all luminosities ranging from $\sim$ 10$^{19}$ W Hz$^{-1}$ to $\sim$ 10$^{29}$ W Hz$^{-1}$. 
This shows that KSRs are found in sources of low as well as high radio luminosities. 
We further note that a small fraction (90/827 $\sim$ 11$\%$) of KSR sources characterized with high ${\theta}_{\rm FIRST}$ ($>$ 2) are 
preferentially found at high radio luminosities {\ie}L$_{\rm 1.4~GHz}$ $>$ 10$^{24}$ W Hz$^{-1}$.
The high radio luminosities (L$_{\rm 1.4~GHz}$) $\sim$ 10$^{24}$ - 10$^{28}$ W Hz$^{-1}$ of these sources imply that they are radio-loud AGN 
wherein L$_{\rm 1.4~GHz}$ $\sim$ 10$^{24.5}$ W Hz$^{-1}$ can be considered as the dividing line between FR-I and FR-II 
radio galaxies \citep{Fanaroff74,Rafter11}.  
Sources with ${\theta}_{\rm FIRST}$ $>$ 2 means that extranuclear flux density is three times of the peak flux 
density which in turn makes them extended-emission-dominated radio sources. 
Indeed, several optically classified Seyfert galaxies are known to be radio-loud AGN with lobe-dominated KSRs (see \cite{Doi12}). 
Moreover, we note that the sources of high radio luminosities (L$_{\rm 1.4~GHz}$ $>$ 10$^{24}$ W Hz$^{-1}$) may have radio galaxy contaminants 
that are either incorrectly classified as Seyfert galaxies or there 
seems to be a population of sources that can be classified as radio galaxies based on its radio properties but optical properties are 
similar to Seyfert galaxies. 
In general, Seyfert galaxies are found to be hosted in late-type disk or lenticular galaxies \citep{Schawinski11}, 
while radio-loud AGN are mainly hosted in early-type galaxies (Best et al. 2005). 
However, in recent times a few examples of radio galaxies hosted in disk galaxies have also been reported
({\eg}NGC 612, \cite{Emonts08}; Speca, \cite{Hota11}; PKS 1814-637, \cite{Morganti11}, J2345-0449, \cite{Bagchi14}). 
The optical and mid-IR properties of host galaxy ISM of these sources often show more in common with Seyfert galaxies 
than they do with radio galaxies \citep{Morganti11}. 
Therefore, a detailed investigation on the properties of host galaxies and ISM of radio powerful Seyfert galaxies is required 
to confirm their nature.
\par
Furthermore, unlike sources with high ${\theta}_{\rm FIRST}$, several sources with high NVSS$-$to$-$FIRST flux density ratios 
(say ${\theta}_{\rm NVSS-FIRST}$ $>$ 1.5) are found at lower radio luminosities. 
A comparison of radio luminosity distributions of simple (${\theta}_{\rm NVSS-FIRST}$ $\leq$ 1.175) and 
complex radio sources (${\theta}_{\rm NVSS-FIRST}$ $\geq$ 1.175) shows that the complex sources indeed have systematically lower 
radio luminosities (with median radio luminosity L$_{\rm 1.4~GHz, FIRST,~median}$ $\sim$ 5.1 $\times$ 10$^{22}$ W Hz$^{-1}$) than simple 
sources (with median radio luminosity L$_{\rm 1.4~GHz, FIRST,~median}$ $\sim$ 4.4 $\times$ 10$^{23}$ W Hz$^{-1}$). 
This implies that low-surface-brightness extended radio emission detected by NVSS is more often present in low luminosity sources.       
Thus, the additional radio emission detected in NVSS may have significant contribution from star$-$formation as AGN are relatively weaker. 

\subsection{Radio loudness}
\begin{figure*}
\includegraphics[angle=0,width=9.2cm]{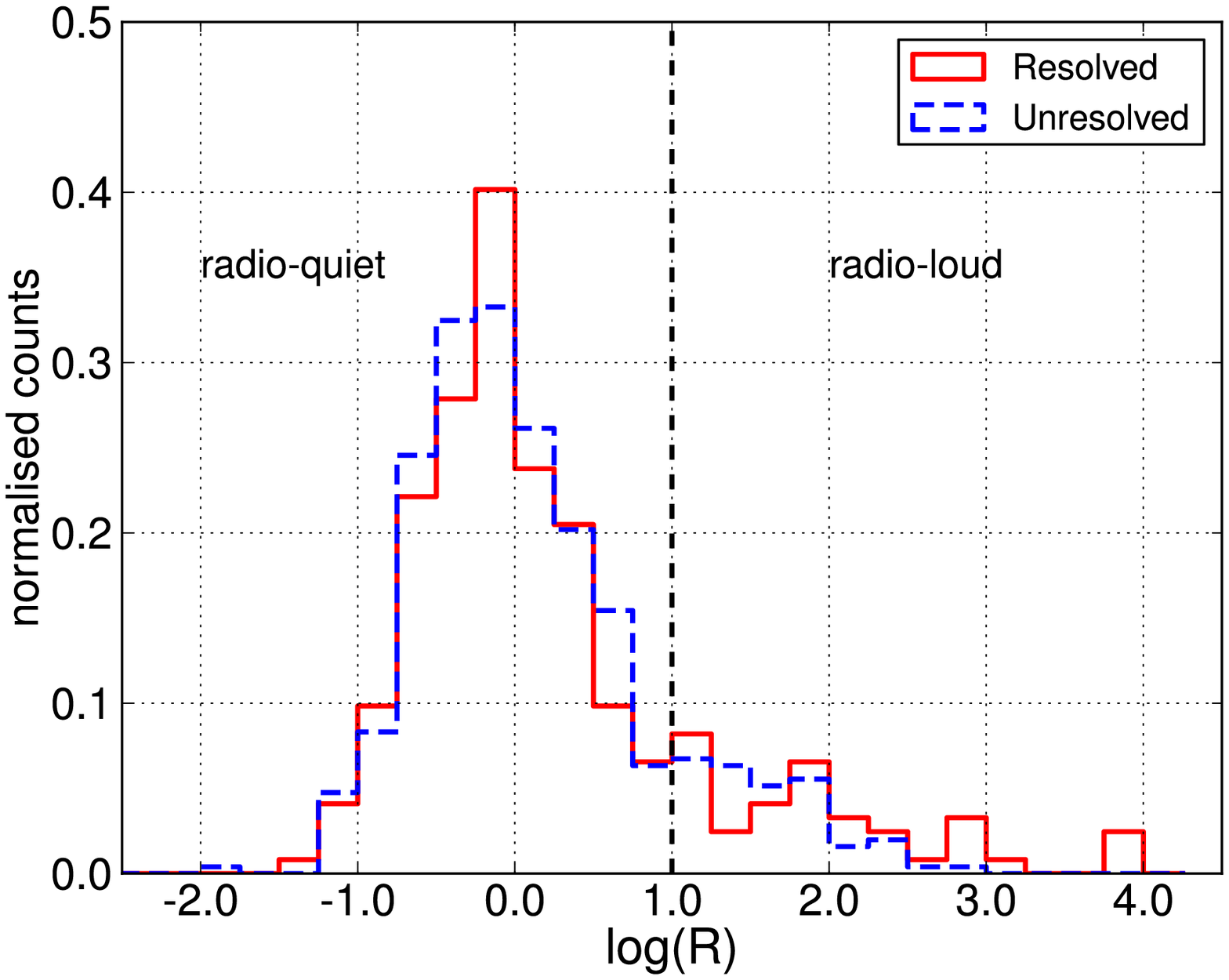}{\includegraphics[angle=0,width=9.2cm]{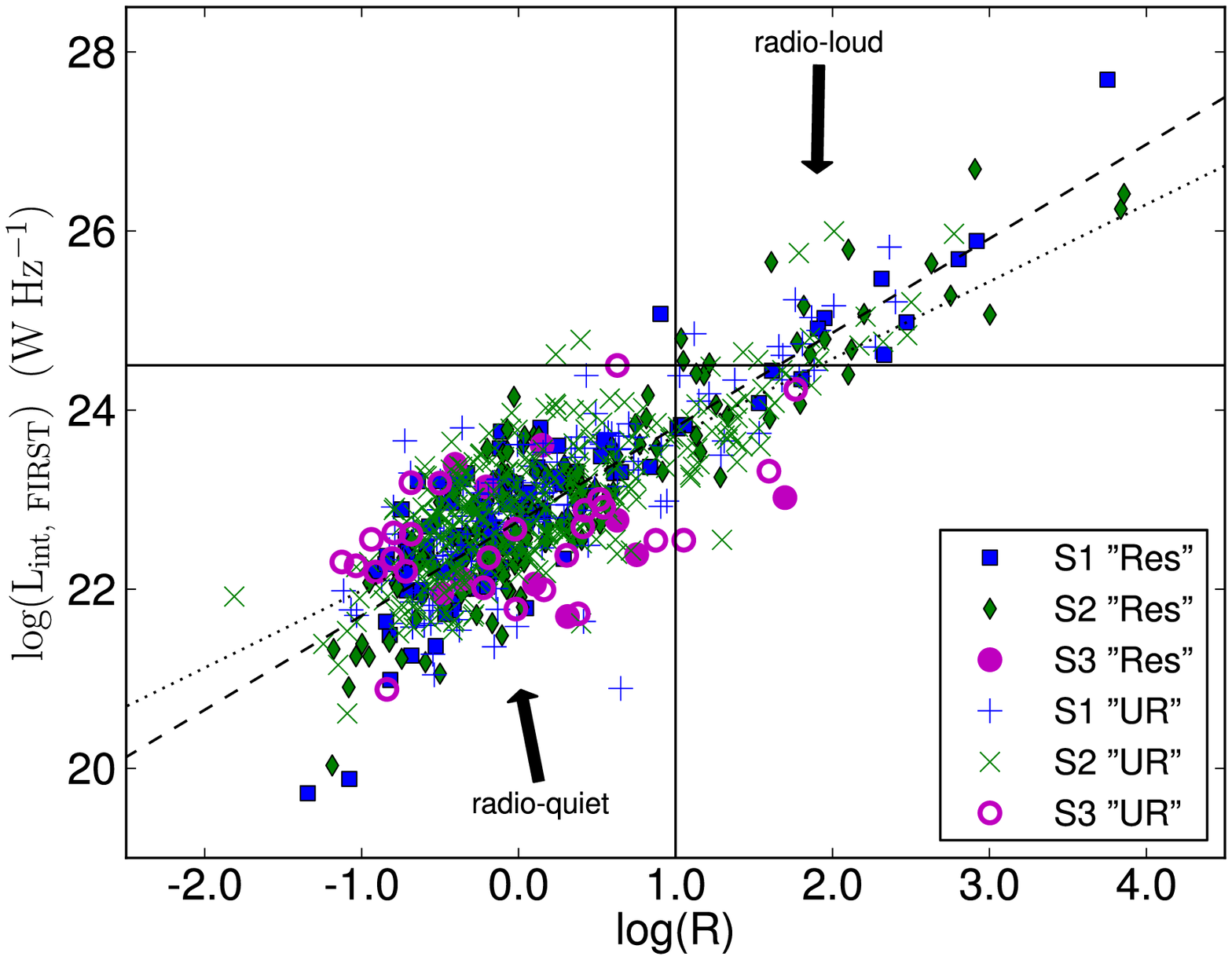}}
\caption{{\it Left :} Distributions of radio$-$loudness parameter for unresolved (${\theta}_{\rm FIRST}$ $\leq$ 1.06) and resolved 
(${\theta}_{\rm FIRST}$ $>$ 1.06) sources. Dotted vertical line is dividing line between radio-quiet and radio-loud population. 
{\it Right :} Radio luminosity versus radio loudness plot. Dashed and dotted lines represent least square regression lines for 
resolved ("Res") sources (log L$_{\rm int,~FIRST}$ $=$ 1.05 logR $+$ 22.76) and unresolved ("UR") sources 
(log L$_{\rm int,~FIRST}$ $=$ 0.86 logR $+$ 22.76), respectively. Solid Vertical line at logR $=$ 1.0 and solid horizontal line at 
log L$_{\rm int,~FIRST}$ = 24.5 W Hz$^{-1}$ represent conventional dividing line between radio-quiet and radio-loud sources. 
Filled and open symbols represent resolved and unresolved radio sources, respectively.} 
\label{fig:RadioLoudness}
\end{figure*}

In general, radio source population is known to exhibit bimodal distribution of radio-loudness parameter (R) 
with dividing line at R $=$ 10 that categories radio sources into radio-quiet (R $<$ 10) and radio-loud (R $\geq$ 10), where R is defined as the 
ratio of monochromatic 5 GHz radio luminosity to the 4400 {\AA} optical luminosity 
(R $=$ $\frac{{{\nu}_{\rm 5 GHz}}{\rm L_{\rm 5 GHz}}}{{\nu}_{\rm 4400{\AA}}{\rm L_{\rm 4400 {\AA}}}}$) \citep{Visnovsky92,Kellermann94}. 
The physical origin of the bimodality is generally attributed to the differences in the 
fundamental parameters of AGN-galaxy system such as the mass and spin of black hole, accretion rate and the Hubble type of 
host galaxy \citep{Laor2000,McLure04,Sikora07,Rafter11}.    
Therefore, the degree of radio-loudness can be used to characterize the nature of AGN. 
We study radio-loudness of our sample sources to investigate if the presence of KSRs is related to radio-loudness. 
We derive 5 GHz radio luminosity (L$_{\rm 5~GHz}$) for our sample sources from 1.4 GHz radio luminosity (L$_{\rm 1.4~GHz}$) using a 
typical radio spectral index of -0.7 for Seyfert galaxies \citep{Morganti99,Singh13}.
The optical continuum luminosity at 4400 {\AA} (L$_{{\lambda}4400 {\AA}}$) is derived from 5100 {\AA} luminosity 
(L$_{{\lambda} 5100 {\AA}}$) using a typical optical spectral index of -0.5 \citep{Sikora07}. 
The 5100 {\AA} luminosity (L$_{{\lambda} 5100 {\AA}}$) is empirically estimated from H$_{{\alpha}{\lambda}6563 {\AA}}$ luminosity as both are 
found to be related to bolometric luminosity \citep{McLure04,Green05}. 
Thus, 4400 {\AA} luminosity (L$_{\rm {\lambda} 4400 {\AA}}$) can be estimated from H$_{{\alpha}{\lambda}6563 {\AA}}$ as : 
L$_{\rm {\lambda} 4400 {\AA}}$ = L$_{\rm {\lambda} 5100 {\AA}}$~$({\frac{4400 {\AA}}{5100 {\AA}}})^{-0.5}$ = (2.59 $\times$ 10$^{43}$) 
$({\frac{\rm {L_{H{\alpha}}}}{10^{42}})}^{0.86}$ erg s$^{-1}$ \citep[see][]{Rafter09}. 
The H$_{{\alpha}{\lambda}6563 {\AA}}$ flux for our sample sources is obtained from SDSS DR10. 
There are only $\sim$ 749/2651 sources with H${\alpha}_{{\lambda}6563 {\AA}}$ flux available from SDSS DR10 and therefore our analysis on 
radio-loudness is limited only to this subsample.
Figure~\ref{fig:RadioLoudness} shows the distribution of radio loudness parameter for extended (KSRs) and compact (non-KSRs) radio 
sources of our sample.
We find that both extended and compact radio sources show similar distributions of radio loudness parameter. 
As expected majority ($\sim$ 85$\%$) of our Seyfert/LINER sample sources are radio-quiet (R $<$ 10), while $\sim$ 15$\%$ sources fall into 
the radio-loud (R $\geq$ 10) category. 
We find that radio loudness parameter is correlated with total radio luminosity for both compact (unresolved) and extended (resolved) sources 
(see Figure~\ref{fig:RadioLoudness}, right panel).
Spearman rank correlation coefficients for unresolved and resolved sources are $\sim$ 0.70 and 0.78, respectively.  
The observed correlation is consistent with previous studies characterizing radio loudness in terms of radio-luminosity \citep[{\eg}][]{Rafter09}. 
It is evident that most of the radio powerful sources (L$_{\rm 1.4~GHz}$ $\geq$ 10$^{24.5}$ W Hz$^{-1}$) are radio-loud, 
while radio weak (L$_{\rm 1.4~GHz}$ $\leq$ 10$^{23}$ W Hz$^{-1}$) are found to be radio-quiet. 
Moreover, a small fraction ($\sim$ 10$\%$) of our sample sources with 1.4 GHz radio luminosity 
(L$_{\rm 1.4~GHz}$) $\sim$ 10$^{23}$ $-$ 10$^{24.5}$ W Hz$^{-1}$ can be classified as radio-intermediate with radio loudness parameter 
(R) $\sim$ 10 $-$ 100. 
The presence of powerful radio-loud sources (R $>$ 100 and L$_{\rm 1.4~GHz}$ $\geq$ 10$^{24.5}$ W Hz$^{-1}$), 
although only a very small fraction ($\sim$ 3.0$\%$), is apparently unexpected as these sources can simply be categorized as FR-II radio galaxies. 
As discussed in the previous section, these sources are either radio galaxy contaminants or there is a population of 
sources in which radio and optical properties give different classification. 
The radio-loudness versus radio luminosity plot for our sample sources suggests that radio-loudness parameter in Seyfert, LINER galaxies 
is distributed in a continuous fashion and overlaps with that of FR-I/FR-II radio galaxies.    
This is consistent with \cite{Kharb14a} results showing the overlap in the radio-loudness of Seyfert galaxies 
and FR-I radio galaxies with $\sim$ 11$\%$ of Seyfert galaxies in their sample classified as radio-loud. 
Using a sample of broad line AGN \cite{Rafter09} also found no clear demarcation between the radio-loud (R $\geq$ 10) and radio-quiet 
(R $<$ 10) sources but instead fill in a more radio-intermediate population in a continuous fashion. 
Finally, we note that the presence of KSRs in Seyfert and LINER galaxies does not seem to have any dependence on the radio-loudness 
as both extended (KSRs) and compact (non-KSRs) sources show similar distribution of radio-loudness parameter.

\subsection{KSR radio power versus AGN power}
A test on the correlation between the strength of kpc-scale radio emission and the AGN power may provide clues 
about the origin of KSRs. 
A positive correlation between the radio power of KSRs and AGN power can be expected if KSRs are powered by AGN. 
We use core radio power, O[III] $\lambda$ 5007 {\AA} luminosity as the proxies for AGN power.
We derive extended emission from total emission subtracted by peak emission 
(S$_{\rm ext}$ = S$_{\rm int, FIRST}$ - S$_{\rm peak, FIRST}$) and FIRST peak flux density is considered as core radio emission.
However, we caution that our estimates of core and extended radio emission (L$_{\rm core}$) using FIRST peak emission are fairly crude as 
the FIRST peak flux density may contain extended emission in high-$z$ sources and in type 1 sources where AGN jets are oriented 
towards observer. 
High resolution observations detecting core radio emission at parsec-scale would give more robust estimates of core radio emission, 
but high resolution radio data are not available for our sample sources. 
To get more accurate estimate of core and extended radio emission we create a subsample of low redshift sources 
(z $\leq$ 0.01) for which FIRST beam of 5$\arcsec$ corresponds to $\leq$ 1.0 kpc 
and thus give better segregation of core and extended emission. 
However, number of such sources is too small with limited span in luminosity space to give a robust statistical result. 
\subsubsection{KSR radio power versus core radio power}
The nuclear radio emission is optically thin to torus dust obscuration and 
also jets in Seyfert and LINER galaxies are not relativistically beamed \citep{Middelberg04,Ulvestad05}.    
Therefore, AGN core radio emission can be considered an isotropic property in Seyfert galaxies that provides an 
orientation-independent measurement of AGN power \citep{Kukula95,Xu99,Thean01}.
To check if extended KSRs emission is related to the AGN core radio power we plot radio luminosity of extended emission versus 
core radio luminosity, 
Figure~\ref{fig:LcoreVsLext} (left panel) shows that the core radio luminosity is positively correlated with the radio luminosity 
of extended emission component. 
Spearman rank correlation test gives correlation coefficient ($\rho$) $=$ 0.94 with the probability of two parameters being uncorrelated 
(p) $<$ 2.2 $\times$ 10$^{-16}$. 
We also performed statistical test on the correlation between core radio flux density (S$_{\rm core}$ = S$_{\rm peak,~FIRST}$) and 
extended radio flux density (S$_{\rm ext}$ = S$_{\rm int,~FIRST}$ - S$_{\rm peak,~FIRST}$). 
Spearman rank correlation test results correlation coefficient ($\rho$) $=$ 0.69 with the probability of two parameters 
being uncorrelated (p) $<$ 1.0 $\times$ 10$^{-16}$. 
This demonstrates that correlation between core radio luminosity and extended radio emission luminosity is not driven by distance effect.
The correlation between core radio luminosity and extended radio emission luminosity can be interpreted as 
more powerful AGN tend to show more powerful KSRs and therefore, inferring that KSRs are likely to be powered by AGN.
Sources with L$_{\rm core}$ $>$ L$_{\rm ext}$ can be considered as core-dominated sources while sources with 
L$_{\rm core}$ $<$ L$_{\rm ext}$ are extended-emission-dominated sources. 
From figure~\ref{fig:LcoreVsLext} it is evident that extended-emission-dominated sources are preferentially found 
at higher radio luminosities (L$_{\rm core}$ $>$ 10$^{23}$ W Hz$^{-1}$), which is also inferred from the 
${\theta}_{\rm FIRST}$ versus L$_{\rm 1.4 GHz, FIRST}$ plot (see Figure~\ref{fig:LFIRSTVsThetaFIRST}).
\subsubsection{KSR radio power versus [O III] $\lambda$ 5007{\AA} line luminosity}
\begin{figure*}
\includegraphics[angle=0,width=9.2cm]{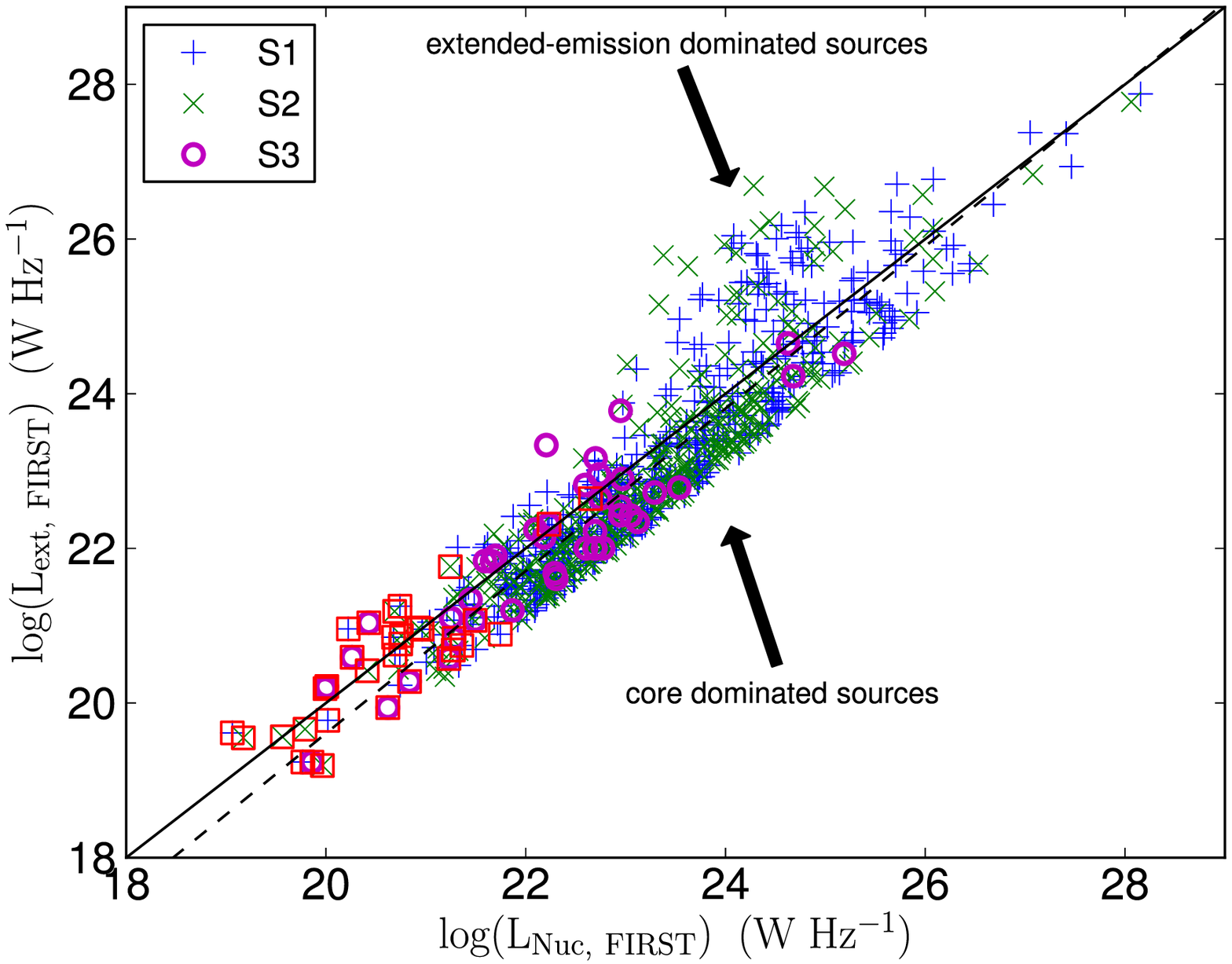}{\includegraphics[angle=0,width=9.2cm]{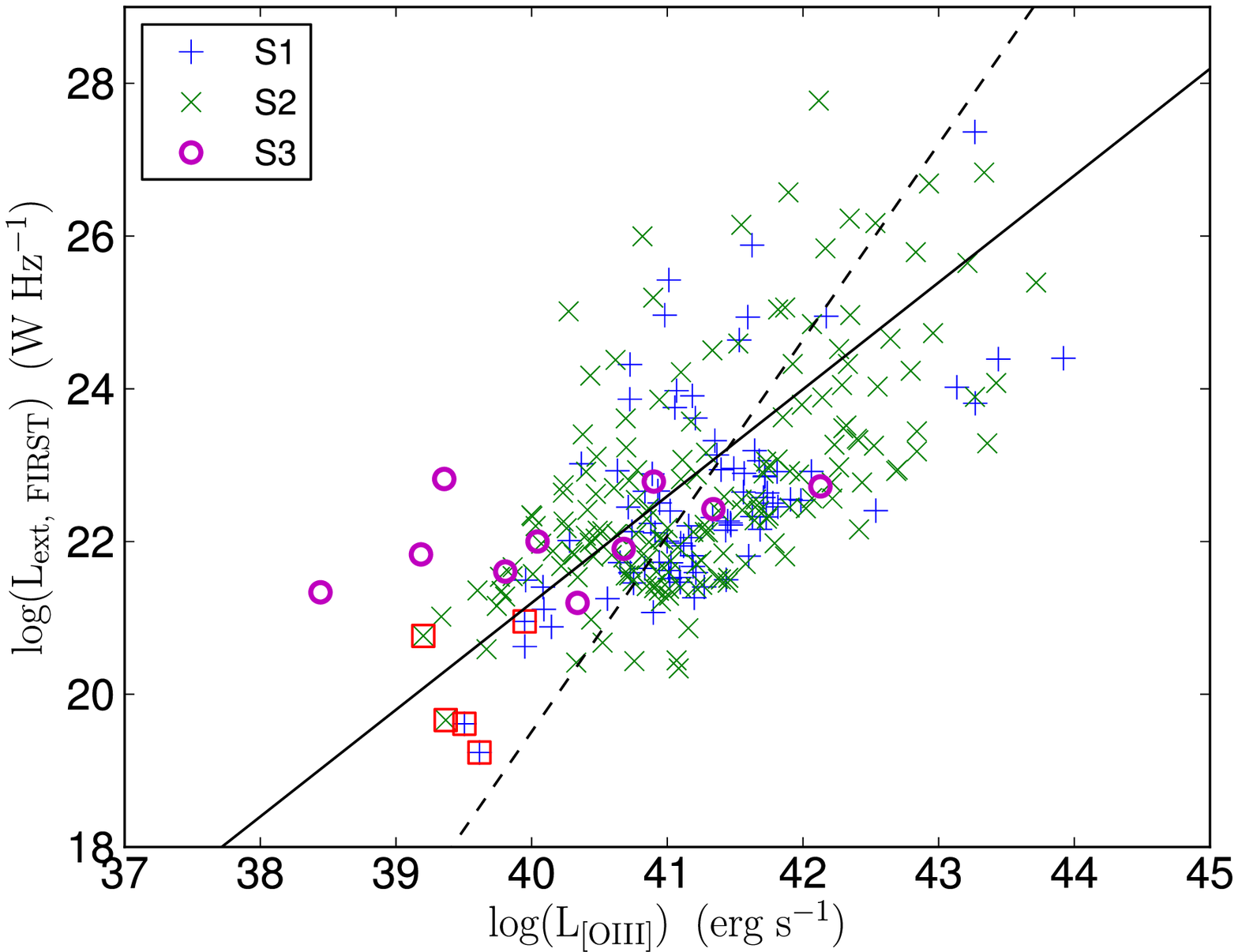}}
\caption{{\it Left :} L$_{\rm ext,~FIRST}$ versus L$_{\rm Nuc,~FIRST}$. 
Solid line represents bisector (L$_{\rm core,~FIRST}$ $=$ L$_{\rm ext,~FIRST}$) while dashed line represents least square regression line 
(L$_{\rm ext, FIRST}$ $=$ 1.05 L$_{\rm core, FIRST}$ $-$ 1.39).
{\it Right :} L$_{\rm ext,~FIRST}$ versus L$_{\rm OIII~{\lambda} 5007{\AA}}$ plot. 
Solid line represents bisector (L$_{\rm [O~III]}$ $=$ 0.69 L$_{\rm ext, FIRST}$ + 25.64) while dashed line represents least square regression line 
(L$_{\rm [O~III]}$ $=$ 0.39 L$_{\rm core, FIRST}$ $+$ 32.39).
Seyfert 1s, Seyfert 2s and LINERs are shown by `+', `$\times$' and circles, respectively. Square symbols represent sources with redshift 
(z) $\leq$ 0.01 for which FIRST beam of $\sim$ 5$^{\arcsec}$.0 corresponds to $\leq$ 1.0 kpc and thereby gives a more robust separation of 
nuclear (core) and extranuclear (extended) emission.}
\label{fig:LcoreVsLext}
\end{figure*}

[O III] $\lambda$ 5007 {\AA} line emission is believed to originate from narrow line region and is found to be correlated with nuclear ionizing
continuum, as well as nuclear X-ray luminosity, and therefore is considered as the proxy for intrinsic AGN power \citep{Nelson95,Heckman05}. 
Also, it is unaffected by the torus obscuration as it originates outside the torus.
Figure~\ref{fig:LcoreVsLext} (right panel) shows an increasing trend of KSRs extended emission radio luminosity (L$_{\rm ext}$) with 
[O III] luminosity, however with a large scatter. 
Spearman rank correlation test yields correlation coefficient ($\rho$) $=$ 0.59 with the probability of two parameters being uncorrelated 
(p) $<$ 1.0 $\times$ 10$^{-16}$. 
The apparent correlations between extended emission (KSRs) radio luminosity (L$_{\rm ext}$) and [O~III] $\lambda$ 5007{\AA} line emission luminosity 
(L$_{\rm [O~III] 5007 {\AA}}$) can be interpreted if extended radio emission is related to AGN power. 
If [O III] $\lambda$ 5007 {\AA} line luminosity is considered as the accurate measure of AGN power then 
the large scatter can be understood if KSRs emission is influence by other factors such as the contamination from star formation, 
presence of circumnuclear starburst and the variation in the type of host galaxy.
\subsection{Ratio of FIR-to-radio flux}
\begin{figure*}
\includegraphics[angle=0,width=9.2cm]{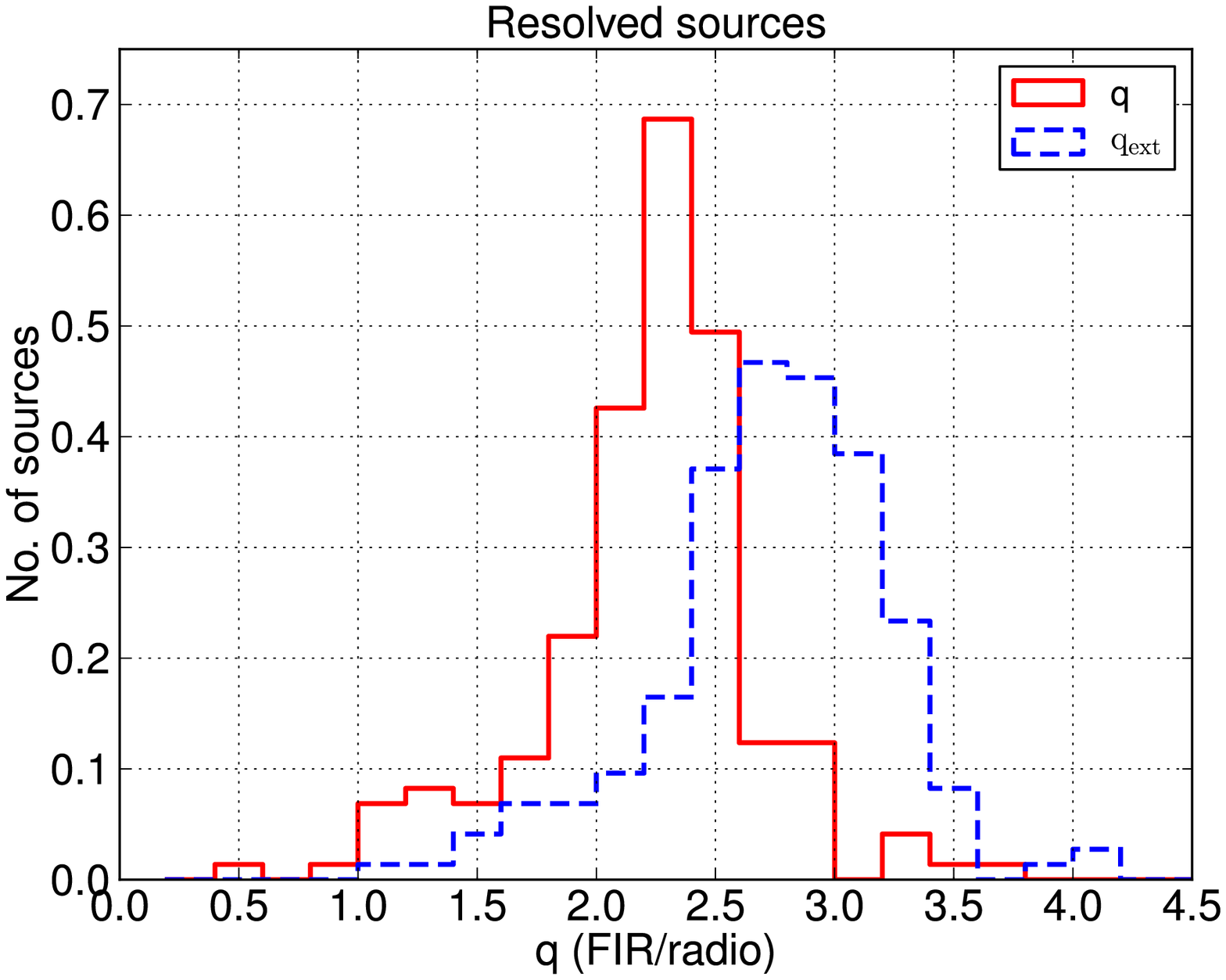}{\includegraphics[angle=0,width=9.2cm]{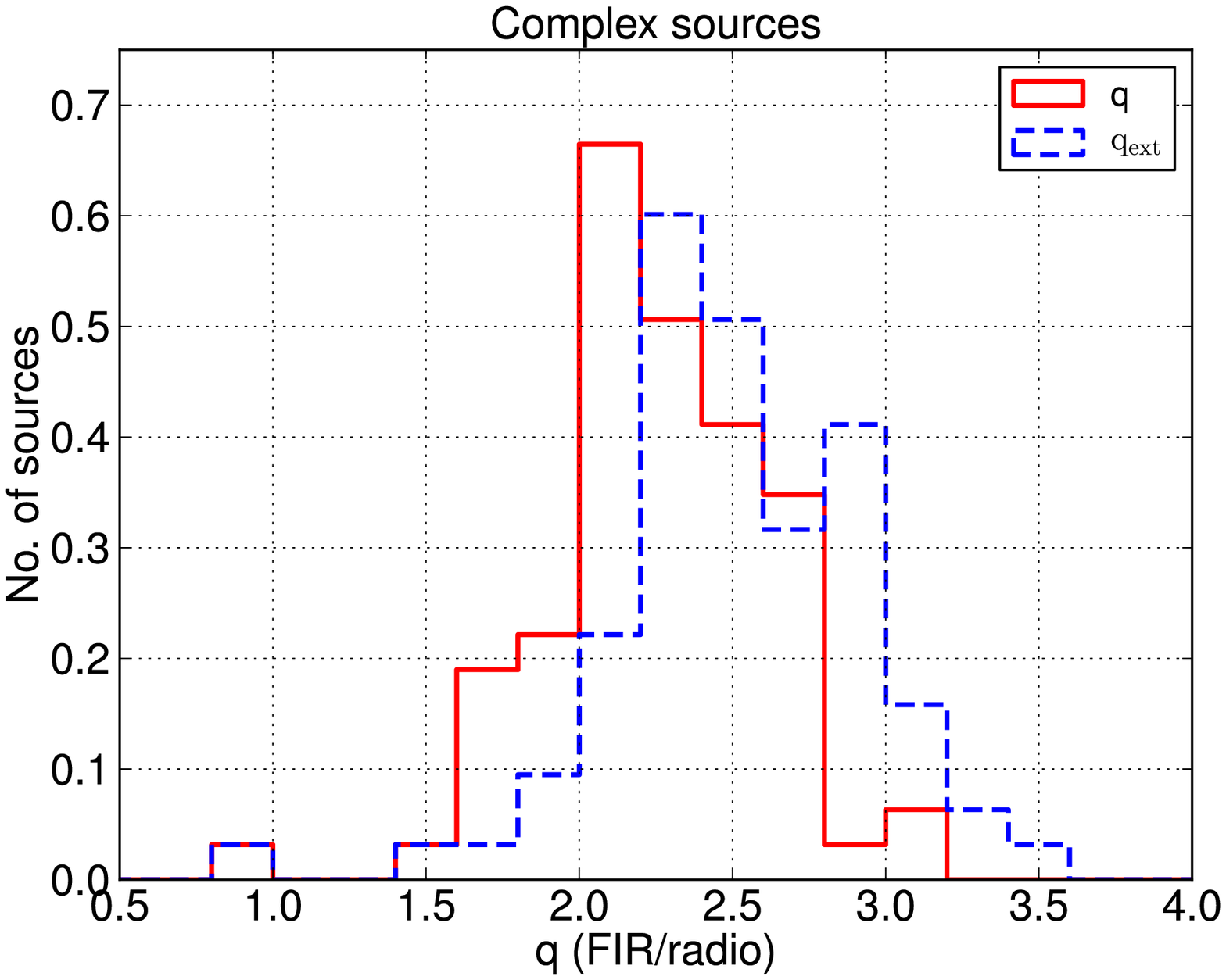}}
\caption{{\it Left :} Distribution of the ratio of FIR to 1.4 GHz radio continuum (q) for resolved radio sources 
(${\theta}_{\rm FIRST}$ $>$ 1.06). {\it Right :} Distribution of the ratio of FIR to 1.4 GHz radio continuum (q) for complex radio sources 
(${\theta}_{\rm FIRST-NVSS}$ $>$ 1.175). 
For resolved and complex sources the parameter q is estimated using total 1.4 GHz radio emission detected in FIRST and  NVSS, respectively. 
Parameter q$_{\rm ext}$ for resolved and complex sources is estimated using only extranuclear radio emission 
(S$_{\rm ext,~FIRST}$ = S$_{\rm int,~FIRST}$ - S$_{\rm peak,~FIRST}$ and S$_{\rm ext,~NVSS}$ = S$_{\rm int,~NVSS}$ - S$_{\rm peak,~FIRST}$) 
detected in FIRST and NVSS, respectively.}
\label{fig:q}
\end{figure*}

Star-forming galaxies are known to display a tight correlation between far-infrared and radio continuum 
emission as both FIR and radio emission are believed to be linked to star-formation activity \citep{Condon92,Kovacs06,Sargent10}. 
However, AGN exhibit radio excess due to additional radio emission from AGN core and jets, 
and tend to deviate from typical radio-FIR correlation \citep{Yun01,Moric10}. 
AGN galaxies can be expected to show tighter radio-FIR correlation if radio emission component from AGN is subtracted out. 
Using a sample of Seyfert galaxies \cite{Baum93} reported that extranuclear radio power derived by subtracting nuclear radio power 
from the total radio power exhibit tighter corelation with the total FIR luminosity.
Following a similar approach we use radio-FIR correlation to examine whether KSRs are powered by AGN or star-formation. 
In Seyfert and LINER galaxies, if extranuclear KSR emission is attributed to star-formation and nuclear emission is solely from AGN core 
then extranuclear radio emission is expected to show tighter correlation with total FIR emission. 
We derive extranuclear KSRs radio emission (S$_{\rm ext}$) by subtracting FIRST peak emission (S$_{\rm peak}$) from the 
NVSS total radio emission (S$_{\rm total}$) and assume that the FIRST peak radio emission is primarily from AGN core. 
\\
We investigate the ratio of FIR to radio continuum (q parameter) for our KSR sample sources using both total and extranuclear radio emission. 
The ratio of FIR to radio continuum is defined as q $=$ log[(FIR/3.75 $\times$ 10$^{12}$ Hz)/S$_{\rm 1.4~GHz}$], where 
FIR $=$ 1.26 $\times$ 10$^{-14}$[2.58S$_{\rm 60~{\mu}m}$ + S$_{\rm 100~{\mu}m}$] W m$^{-2}$ \citep{Helou85}. 
The FIR fluxes at 60~$\mu$m and 100~$\mu$m are obtained from InfraRed Astronomical Survey 
(IRAS\footnote{http://irsa.ipac.caltech.edu/Missions/iras.html}) and are in Jy units. 
Using IRAS observations we find FIR counterparts of only 217/776 $\sim$ 28$\%$ KSR sources and therefore our 
investigation on radio-FIR correlation is limited to FIR detected subsample.
Fig~\ref{fig:q} shows the distributions of the ratio of FIR to radio (q parameter) for resolved (${\theta}_{\rm FIRST}$ $>$ 1.06) and 
complex (${\theta}_{\rm FIRST-NVSS}$ $>$ 1.175) KSR sources. 
We find that both resolved and complex sources exhibit wide range of q values over 0.5 to 4.0 with 
median values $\sim$ 2.27 and 2.21, respectively, with standard deviations 0.56 and $\sim$ 0.37, respectively. 
The average value of q parameter for star-forming galaxies in the local universe is found to be (q) $\sim$ 2.3, and it decreases for 
AGN-bearing galaxies \citep{Sargent10}. 
The lower median value and wide distribution of q parameter for our (resolved and complex) KSR sources is indicative of the presence of AGN.
\\  
We also estimate the ratio of FIR$-$to$-$radio (q$_{\rm ext}$) using extranuclear flux density for resolved and complex KSR sources.
The extranuclear radio emission for KSR sources is derived by subtracting FIRST peak flux density from total FIRST and NVSS flux 
density, respectively ({\ie}S$_{\rm ext~FIRST}$ = S$_{\rm int,~FIRST}$ - S$_{\rm peak,~FIRST}$ 
and {\ie}S$_{\rm ext~NVSS}$ = S$_{\rm int,~NVSS}$ - S$_{\rm peak,~FIRST}$). 
We note that that both resolved and complex KSR sources show an even wider distribution of q$_{\rm ext}$ spanning over 0.5 to 4.5 with 
median values $\sim$ 2.77 and 2.46 respectively and standard deviation $\sim$ 0.58 and 0.41, respectively. 
The wider distribution of q$_{\rm ext}$ compared to q suggests that extranuclear KSR radio emission is likely to be powered by AGN \cite[{\eg}][]{Roy98}. 
The higher median values of q$_{\rm ext}$ can be understood if extranuclear total radio emission is underestimated as the 
FIRST peak emission can have significant contribution from star-formation. 
The dusty torus around AGN may also contribute significantly to total FIR emission and can result in a higher value of q$_{\rm ext}$ parameter.

\subsection{Mid-IR colors}
\begin{figure}
\includegraphics[angle=0,width=9.2cm]{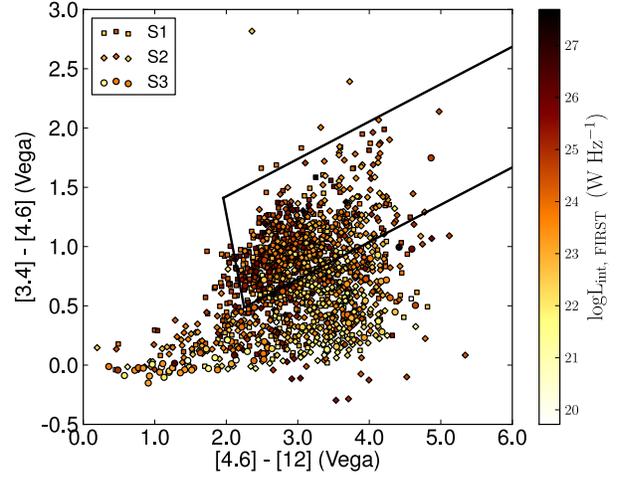}
\caption{ WISE mid-IR colors for KSR sources. 
Solid lines show the AGN wedge defined as : W1 $-$ W2 $>$ 0.315 $\times$ (W2 $-$ W3) $-$ 0.222 and 
W1 $-$ W2 $<$ 0.315 $\times$ (W2 $-$ W3) + 0.796 and W1 - W2 $=$ -3.172 $\times$ (W2 − W3) + 7.624) \citep{Mateos12}.}
\label{fig:IRColors}
\end{figure}

IR colors can be used to measure the relative contribution of a starburst or AGN to the total bolometric luminosity, with 
redder (cooler) colors characterizing the dominance of starburst \citep{Rush93,Dopita98,Lacy04,Stern12}. 
We use mid-IR color-color diagram to check if KSR sources are dominated by starburst or AGN using 
{\em Wide-field Infrared Survey Explorer (WISE)\footnote{http://irsa.ipac.caltech.edu/Missions/wise.html}} data. 
WISE covers all sky in four mid-IR bands {\ie}W1 [3.4 $\mu$m], W2 [4.6 $\mu$m], W3 [12 $\mu$m], and W4 [22 $\mu$m]) 
with an angular resolution of 6.1, 6.4, 6.5, and 12 arcsec, respectively.
WISE counterparts of our sources are found using a search radius of 2$^{\arcsec}$.0 around the optical positions. 
Only 752/827 sources (91$\%$) with extended radio emission have WISE counterparts.
\\
In the literature, several mid-IR color-color diagnostics have been developed for separating AGN and star-forming normal galaxies 
\citep[{\eg}][]{Lacy04,Stern05} and WISE colors have been used to identify AGN \citep[{\eg}][]{Stern12,Mateos12}). 
We use WISE color-color diagrams of our sample sources, where the AGN selection wedge is defined as : 
W1 $-$ W2 $\geq$ 0.315 $\times$ (W2 − W3) $-$ 0.222 and W1 $-$ W2 $\geq$ 0.315 $\times$ (W2 − W3) + 0.796 and 
W1 - W2 $=$ -3.172 $\times$ (W2 − W3) + 7.624 \citep[see][]{Mateos12}.
Using 22 $\mu$m WISE band is not much useful due to its significantly shallower depth in comparison to the first three bands and also 
star formation contribution is likely to increase at 22 $\mu$m \citep{Mateos12}. 
Figure~\ref{fig:IRColors} shows that KSR sources have a wide range of IR colors with majority of them falling into AGN wedge.
We note that, in general, sources within the AGN selection wedge are radio powerful sources while sources 
falling outside the AGN wedge are of relatively lower radio luminosities. 
This can be understood if mid-IR emission is dominated by AGN in radio powerful sources (L$_{\rm 1.4 GHz}$ $\geq$ 10$^{24}$ W Hz$^{-1}$), while
in sources with relatively lower radio luminosity mid-IR emission apparently have greater contribution from dust heated by stars.
Therefore, KSRs in radio luminous sources are likely to be powered by AGN. On the other hand, in sources of relatively low radio luminosities, 
KSRs can have contribution from starformation. 
Moreover, we caution that the WISE mid-IR colors cuts should be used with care as these color cuts may be biased against 
heavily obscured type-2 AGN. 
Also, there are suggestions that radio selected AGN may have a different accretion mode, {\ie}radiatively inefficient (radio mode), 
and may not strictly follow the mid-IR color selection criteria \citep{Hardcastle07,Griffith10}.

\section{Conclusions}
We have studied the prevalence and nature of KSRs in Seyfert and LINER galaxies using 1.4 GHz VLA FIRST and NVSS observations.
Our sample is the largest sample hitherto used for KSR studies and consists of 2651 sources detected in FIRST. Of these, 1737 sources also 
have NVSS counterparts. 
Similar redshift distributions of Seyfert type 1s and type 2s suggests 
that our sample is not affected by biases introduced by dust obscuration.  
Conclusions of our study are outlined as below. 

\begin{enumerate}
\item
Using a concentration parameter based on the ratio of total$-$to$-$peak flux density 
${\theta}_{\rm FIRST}$ $=$ ${\rm (S_{\rm int,~FIRST}/S_{\rm peak,~FIRST})^{1/2}}$, 
we find $\geq$ 30$\%$ KSR sources ({\ie}sources with radio structures larger than 1.0 kpc) in the FIRST detected sample. 
The comparison of NVSS to FIRST flux densities help us in detecting faint extended radio emission that is resolved out in FIRST observations and 
we obtain $\geq$ 42$\%$ KSR sources in FIRST-NVSS subsample. 
This fraction of KSR sources is only a lower limit owing to the combined effects of projection, resolution and sensitivity. 
The comparison of the fraction of KSRs detected in our sample with the previous studies demonstrates the importance 
of low-frequency observations with relatively lower resolution in detecting KSRs efficiently. 
\item
There are a total of 180 Seyfert and LINER galaxies in our sample that exhibit multicomponent KSRs in FIRST observations.
We find that these sources display varying morphologies {\eg}linear, S-shaped and diffuse, which is consistent with 
the previous studies based on sensitive targeted observations of small samples.  
\item 
We find that KSR sources are distributed across all redshifts, luminosities and radio-loudness.  
We show that the presence of KSRs in Seyfert and LINER galaxies does not seem to have any dependence on the radio luminosity and radio-loudness.
A small fraction ($\sim$ 10$\%$ - 15$\%$) of KSR sources characterized with high total$-$to$-$peak flux density ratio 
and high radio luminosity (L$_{\rm 1.4~GHz}$ $>$ 10$^{24}$ W~Hz$^{-1}$) are radio$-$loud.
These sources may constitute a population of radio powerful Seyfert and LINER galaxies with radio properties overlapping with radio galaxies. 
\item
We find that the extranuclear radio power in KSR sources is positively correlated with the core radio power and [O~III] $\lambda$ 5007{\AA} line 
luminosity. This can be interpreted as KSRs are powered by AGN rather than star$-$formation.
\item
The ratio of FIR$-$to$-$radio continuum (q parameter) for KSR sources shows a wide distribution spanning over 0.5 to 4.0 with 
median value lower than that for typical starforming galaxies and thus indicating the dominance of AGN radio emission. 
The ratio of FIR$-$to$-$radio continuum (q$_{\rm ext}$) using extranuclear radio flux density (derived by subtracting FIRST peak flux density 
from NVSS total flux density) shows an even wider distribution suggesting that the extranuclear KSR radio emission is likely to be originated from AGN. 
\item
The mid-IR color-color diagram for KSR sources shows that the mid-IR emission in radio powerful sources tends to be dominated by AGN, 
while in sources with lower radio luminosities mid-IR emission apparently have greater contribution from dust heated by stars.
This can be understood if KSRs in radio luminous sources are primarily powered by AGN, while in sources with relatively low
radio luminosities KSRs may have substantial contribution from star$-$formation.

\end{enumerate}

\section*{Acknowledgments}
We acknowledge support from the Indo-French Center for the Promotion of Advanced Research (Centre Franco-Indien pour la
Promotion de la Recherche Avance) under program no. 4404-3. 
%
%
This research has made use of the NASA/IPAC Extragalactic Database (NED) which is operated by the Jet Propulsion Laboratory, California 
Institute of Technology, under contract with the National Aeronautics and Space Administration. 
This publication makes use of data products from the Wide-field Infrared Survey Explorer, which is a joint project of the University of California, 
Los Angeles, and the Jet Propulsion Laboratory/California Institute of Technology, funded by the National Aeronautics and Space Administration.
\\
Funding for SDSS-III has been provided by the Alfred P. Sloan Foundation, the Participating Institutions, the National Science Foundation, 
and the U.S. Department of Energy Office of Science. The SDSS-III web site is http://www.sdss3.org/.
SDSS-III is managed by the Astrophysical Research Consortium for the Participating Institutions of the SDSS-III Collaboration 
including the University of Arizona, the Brazilian Participation Group, Brookhaven National Laboratory, Carnegie Mellon University, 
University of Florida, the French Participation Group, the German Participation Group, Harvard University, the Instituto de Astrofisica de Canarias, 
the Michigan State/Notre Dame/JINA Participation Group, Johns Hopkins University, Lawrence Berkeley National Laboratory, 
Max Planck Institute for Astrophysics, Max Planck Institute for Extraterrestrial Physics, New Mexico State University, 
New York University, Ohio State University, Pennsylvania State University, University of Portsmouth, Princeton University, 
the Spanish Participation Group, University of Tokyo, University of Utah, Vanderbilt University, University of Virginia, 
University of Washington, and Yale University. 

\bibliographystyle{mn2e}
\bibliography{firstnvsspaper}

\appendix

\section[]{FIRST and NVSS observed parameters for our sample LLAGN}
Table A1 lists the FIRST and NVSS parameters for the first 10 sources of our sample.

\begin{table*}
\begin{sideways}
\begin{minipage}{22.5cm}
\caption{FIRST and NVSS counterpart parameters of Seyfert and LINER galaxies}
\begin{tabular}{@{}cccccccccccccccc@{}}
\hline
 \multicolumn{5}{c}{VV10 parameters} &  \multicolumn{5}{c}{FIRST parameters}  &  \multicolumn{4}{c}{NVSS parameters} &    &     \\ \hline
RA         & Dec   &   z     & AGN  & Mv    &  S$_{\rm p}$ & S$_{\rm int}$ & size (max.$\times$ min.)  & PA & R$_{\rm off}^{\rm FIRST}$  & S$_{\rm int}$ & size (max.$\times$ min.) & PA & R$_{\rm off}^{\rm NVSS}$ & ${\theta}_{\rm FIRST}$ & ${\theta}_{\rm NVSS-FIRST}$ \\   
           &       &         & type &       &  (mJy/b)      &  (mJy)       &  (arcsec$^{2}$)  & (deg)  &   ($\arcsec$) &   (mJy)        &     (arcsec$^{2}$)    & (deg) & ($\arcsec$)  &     &    \\ 
   (1)     &   (2)     &  (3)  & (4)&  (5)  &  (6)  &  (7)  &     (8)            &  (9)  &  (10)&  (11) &      (12)            &  (13) & (14)&  (15)&  (16)\\  \hline
00 02 49.1 & +00 45 04 & 0.087 & S1 & 18.23 & 2.36  & 2.36  & 6.26 $\times$ 5.44 & 19.3  & 0.7  & 2.36  & 45.01 $\times$ 45.01 & 19.8  & 9.2 & 1.00 & 1.00 \\
00 03 10.1 & +04 44 55 & 0.058 & S2 & 17.00 & 5.52  & 5.52  & 6.23 $\times$ 5.37 & 175.7 & 1.7  & 6.09  & 51.39 $\times$ 45.75 & 154.8 & 1.3 & 1.00 & 1.07  \\
00 03 18.2 & +00 48 44 & 0.139 & S2 & 19.48 & 3.66  & 3.71  & 6.45 $\times$ 5.43 & 13.6  & 0.4  & 3.66  & 58.50 $\times$ 45.97 & 32.9  & 6.7 & 1.01 & 1.00 \\
00 03 51.9 & -01 01 42 & 0.269 & S2 & 19.92 & 8.45  & 12.31 & 8.76 $\times$ 5.74 & 178.1 & 1.7  & 12.56 & 48.73 $\times$ 45.01 & 168.4 & 3.5 & 1.21 & 1.01 \\
00 04 58.6 & +11 42 04 & 0.074 & S2 & 17.00 & 13.14 & 13.90 & 6.59 $\times$ 5.55 & 172.6 & 2.5  & 30.26 & 46.63 $\times$ 45.99 & -81.9 & 0.4 & 1.03 & 1.48 \\
00 06 28.5 & -03 42 57 & 0.021 & S3 & 13.90 & 2.37  & 2.55  & 6.17 $\times$ 6.02 & 179.1 & 2.2  & 2.55  & 58.17 $\times$ 46.91 & -2.7  & 8.9 & 1.04 & 1.00 \\
00 09 08.0 & +14 27 55 & 0.040 & S2 & 15.31 & 8.60  & 8.69  & 6.35 $\times$ 5.50 & 178.8 & 1.6  & 11.01 & 47.48 $\times$ 46.36 & 80.9  & 1.5 & 1.01 & 1.13 \\
00 09 11.5 & -00 36 55 & 0.073 & S2 & 18.27 & 38.94 & 39.67 & 6.47 $\times$ 5.44 & 1.2   & 1.6  & 39.67 & 46.78 $\times$ 46.48 & -52.1 & 1.9 & 1.01 & 1.00 \\
00 12 26.8 & -00 48 19 & 0.073 & S1 & 18.16 & 5.03  & 5.30  & 6.72 $\times$ 5.42 & 176.4 & 0.7  & 5.30  & 55.72 $\times$ 48.59 & -173.1& 6.4 & 1.03 & 1.00 \\
00 19 38.9 & -09 40 26 & 0.085 & S3 & 15.90 & 3.11  & 3.11  & 5.84 $\times$ 5.10 & 17.3  & 0.8  & 4.29  & 46.76 $\times$ 45.01 & -63.2 & 2.6 & 1.00 & 1.27 \\
00 20 34.7 & -00 28 14 & 0.072 & S2 & 18.85 & 9.44  & 9.53  & 6.50 $\times$ 5.37 & 171.1 & 0.6  & 10.22 & 50.78 $\times$ 45.29 & 27.2  & 2.3 & 1.01 & 1.04 \\
\hline
\end{tabular}
Notes: First 10 entries of our cross-matched catalog of VV10 and FIRST and NVSS for our sample of Seyfert and LINER galaxies.\\
Columns (1) - (2) : RA and DEC of the optical positions from VV10 catalog; Column (3) : redshift ($z$); Column (4) : Seyfert type {\ie}Seyfert type 1s, 
type 2s and LINERs are designated as S1, S2 and S3, respectively. Column (5) : observed B-band optical magnitude; Column (6) : FIRST peak flux 
density (S$_{\rm p}$) in mJy; Column (7) : FIRST integrated flux density in mJy; Column (8) : FIRST fitted source size in arcsec; 
Column (9) : Position angle of FIRST source; Column (10) : Offset of FIRST counterpart from optical position (in arcsec); 
Column (11) : NVSS total flux density in mJy; Column (12) : NVSS fitted source size; Column (13) : Position angle of NVSS source; 
Column (14) : Offset of NVSS position from optical position; Column (15) : ${\theta}_{\rm FIRST}$ $=$ ${\rm (S_{\rm int,~FIRST}/S_{\rm peak,~FIRST})^{1/2}}$; 
Column (16) : ${\theta}_{\rm NVSS-FIRST}$ $=$ ${\rm (S_{\rm int,~NVSS}/S_{\rm int,~FIRST})^{1/2}}$.
\end{minipage}
\end{sideways}
\end{table*}

\bsp

\label{lastpage}

\end{document}